\documentclass[twocolumn,showpacs,superscriptaddress,preprintnumbers,amsmath,amssymb]{revtex4-1}
\usepackage{graphicx}
\usepackage{dcolumn}
\usepackage{appendix}
\usepackage{bm}

\usepackage[usenames,dvipsnames]{xcolor}
\usepackage{subfigure}
\usepackage{bbm}
\usepackage{enumerate}
\usepackage[
  bookmarks=true,
  colorlinks,
  linkcolor=black,
  urlcolor=black,
  citecolor=black,
  plainpages=false,
  pdfpagelabels,
  final,
  breaklinks=true
]{hyperref}

\usepackage{titlesec}
\usepackage{array}
\usepackage{dsfont}
\usepackage{physics}
\usepackage{mathtools}

\begin{document}

\title{Quantum state engineering of light using intensity measurements and post-selection}

\author{J.~Rivera-Dean}
\email{javier.rivera@icfo.eu}
\affiliation{ICFO -- Institut de Ciencies Fotoniques, The Barcelona Institute of Science and Technology, Castelldefels (Barcelona) 08860, Spain.}

\author{Th. Lamprou}
 \affiliation{Foundation for Research and Technology-Hellas, Institute of Electronic Structure \& Laser, GR-7001 Heraklion (Crete), Greece}

\author{E. Pisanty}
\affiliation{Attosecond Quantum Physics Laboratory, King’s College London, London WC2R 2LS, UK}

\author{M.~F.~Ciappina}
\affiliation{Department of Physics, Guangdong Technion - Israel Institute of Technology, 241 Daxue Road, Shantou, Guangdong, China, 515063}
\affiliation{Technion - Israel Institute of Technology, Haifa, 32000, Israel}
\affiliation{Guangdong Provincial Key Laboratory of Materials and Technologies for Energy Conversion, Guangdong Technion - Israel Institute of Technology, 241 Daxue Road, Shantou, Guangdong, China, 515063}

\author{P. Tzallas}
\affiliation{Foundation for Research and Technology-Hellas, Institute of Electronic Structure \& Laser, GR-7001 Heraklion (Crete), Greece}
\affiliation{ELI-ALPS, ELI-Hu Non-Profit Ltd., Dugonics tér 13, H-6720 Szeged, Hungary}

\author{M. Lewenstein}
\affiliation{ICFO -- Institut de Ciencies Fotoniques, The Barcelona Institute of Science and Technology, Castelldefels (Barcelona) 08860, Spain.}
\affiliation{ICREA, Pg. Llu\'{\i}s Companys 23, 08010 Barcelona, Spain}

\author{P. Stammer}
\email{philipp.stammer@icfo.eu}
\affiliation{ICFO -- Institut de Ciencies Fotoniques, The Barcelona Institute of Science and Technology, Castelldefels (Barcelona) 08860, Spain.}
\affiliation{Atominstitut, Technische Universit\"{a}t Wien, 1020 Vienna, Austria}

\date{\today}

\begin{abstract}

Quantum state engineering of light is of great interest for quantum technologies, particularly generating non-classical states of light, and is often studied through conditioning approaches. 
Recently, we demonstrated that state engineering approaches can be applied in intense laser-atom interactions to obtain measurement statistics which resemble optical “cat” states by using intensity measurements and classical post-selection of the measurement data.
Post-processing of the finite-size sampled data set allows to select specific events, here the processes which are energy conserving, corresponding to measurement statistics of non-classical states of light. 
However, to fully realize the potential of this method for quantum state engineering, it is crucial to thoroughly investigate the role of the involved measurements and the finite-size nature of the post-selection scheme.
We illustrate this by analyzing the post-selection approach recently developed for high harmonic generation, which enables the generation of optical cat states bright enough to induce non-linear phenomena. These findings provide significant guidance for quantum light engineering and the generation of intense optical cat states with potential applications in non-linear optics and quantum science.

\end{abstract}

\maketitle

\section{Introduction}

The last five years witnessed a renewed interest in the intersection of quantum optics and strong-field physics~\cite{lewenstein_attosecond_2022,bhattacharya_stronglaserfield_2023, cruz2024quantum}. Historically, this interest can be traced back to the 1990s~\cite{guo_quantum_1988,guo_scattering_1989,aberg_scattering-theoretical_1991,guo_multiphoton_1992}, when most studies focused on providing more complete quantum electrodynamic descriptions of strong-field processes~\cite{gao_nonperturbative_2000,fu_interrelation_2001,wang_frequency-domain_2007,wang_frequency-domain_2012} or describing novel mechanisms responsible for these observations~\cite{gao_quantum_1998,chen_comment_2000,eden_eden_2000}. However, the recent growth in investigation has been driven by a different question: the quantum properties of the driving field or the harmonic field modes after the process of high harmonic generation (HHG), a highly non-linear process in which the photons of an intense infrared (IR) driving field are up-converted into radiation spanning from the near-infrared to the extreme ultraviolet (XUV) regime~\cite{burnett_harmonic_1977,mcpherson_studies_1987,ferray_multiple-harmonic_1988}. 
This renewed attention has particularly focused on generating non-classical states of light with unprecedented intensity and frequency regimes~\cite{lewenstein2021generation,rivera-dean_strong_2022,stammer_high_2022, stammer_theory_2022, stammer_quantum_2023,lamprou_nonlinear_2023, pizzi2023light, tzur_generation_2023, yi2024generation, gonoskov2024nonclassical, gombkotHo2024parametric, stammer2024entanglement}. 
Additionally, the use of non-classical states of light to drive strong-field processes can influence the electron dynamics~\cite{even_tzur_photon-statistics_2023,even_tzur_motion_2024}, alter the HHG spectrum~\cite{gorlach_high-harmonic_2023} and leave their fingerprints on the state of the light after the interaction~\cite{tzur_generation_2023, lemieux_photon_2024, rasputnyi_high_2024, stammer2024absence}.

It was first demonstrated in Ref.~\cite{gonoskov_quantum_2016} that the process of HHG in atomic systems distinctly alters the photon statistics of the driving field after the interaction. This theoretical insight was later experimentally verified in Ref.~\cite{tsatrafyllis2017high} for atomic systems and extended to semiconductor materials in Ref.~\cite{tsatrafyllis_quantum_2019}. These pioneering experimental studies introduced the idea of post-selection on HHG events by measuring the IR field after the interaction alongside with the generated harmonics~\cite{gonoskov_quantum_2016,tsatrafyllis2017high,tsatrafyllis_quantum_2019}. These post-selection schemes were instrumental for the first observation of non-classical signatures in the IR field after the HHG process, characterized by small negative regions in their Wigner function representation~\cite{lewenstein2021generation}, which were subsequently enhanced in Ref.~\cite{rivera-dean_strong_2022}. Interestingly, in the absence of post-selection protocols, the IR state remained a classical Gaussian state, highlighting the essential role of post-selection in leveraging strong-field physics for engineering non-classical states of light~\cite{stammer_quantum_2023}.

Despite the experimental progress, theoretical explanations are so far based on a phenomenological approach. Although the non-trivial Wigner function of the experiment was observed in the measurement statistics, and the theoretical model is in good agreement with the recorded data, further explanation of the post-selection approach is crucial. 
This includes highlighting the key ingredients of the post-selection scheme: how energy conservation is handled in the post processing of the data, and how the relationship between the measured photocurrent and energy is maintained.
Understanding the role of the shot-to-shot field fluctuations is essential to use the full potential of this scheme for quantum state engineering of light using HHG.

This work addresses the fundamental question of the role of fine-size sampling in strong field post-selection experiments. In order to address this question we investigate the case study of generating massive quantum states of light using post-selection methods in HHG. 
In this paper we take into account the full experimental technicalities, and demonstrate a comprehensive theoretical description of engineering high-quality intense quantum light states using intense laser sources and post-selection approaches after HHG. We examine how the role of energy conservation in the HHG process is properly taken into account in the post-selection, and study its influence on the non-classical features of the Wigner function (cf. \cite{stammer_energy_2024}). We compare these findings with the predictions of previous theoretical approaches and experimental results. Through numerical sampling, we also study the influence of the finite nature of quantum tomography methods used in experimental reconstructions of Wigner functions~\cite{lewenstein2021generation,rivera-dean_strong_2022}, and determine its effects on the observed non-classical features. With these, we provide a more elaborate comparison with the experimental results from Refs.~\cite{lewenstein2021generation,rivera-dean_strong_2022}.
We adopt a didactical approach towards the problem of quantum state tomography with limited data from the numerical, or experimental, sampling particularly addressing the strong field community. 
Thus, this work presents a detailed theoretical work-in-progress story for post-selection schemes in high harmonics generation experiments. 
We provide a clear picture of what the method does and does not demonstrate. We discuss in detail the key steps to create non-classical light depicting non-trivial Wigner functions. This will ultimately help to enhance the fidelity of the practical measurements and assess the potential for additional applications of this method.
Furthermore, going beyond the process of HHG, the presented scheme applies to all parametric light generation processes, in which energy conservation holds.

We emphasize, that the present manuscript does not aim to provide a rigorous theory for the nonlinear interaction leading to the generation of high harmonics. The aim is to provide a numerical sampling experiment based on post-selection techniques from the generated data set using energy conservation. This is in tight connection with the experiment as every real quantum experiment is a sampling experiment.

The remaining manuscript is organized in the following way. In Sec.~\ref{sec:protocol_introduction}, we describe the post-selection protocol in detail, and the conditions under which it leads to the creation of non-classical states of light depicting non-trivial Wigner functions. This is followed by Sec.~\ref{sec:comparison}, where the previously introduced scheme is compared to analytical optical cat states and the experimental generation of non-trivial Wigner functions obtained in Ref.~\cite{lewenstein2021generation, rivera-dean_strong_2022}. For the didactical purpose of the manuscript we include a discussion on imperfections and errors in Sec.~\ref{sec:limitations}. We conclude the manuscript with Sec.~\ref{sec:conclusion}.

\section{\label{sec:protocol_introduction}Post-selection process for creating non-classical light states in HHG}

In this section, we first introduce the platform of quantum optical HHG in Sec.~\ref{sec:stating_problem}, and subsequently introduce the specific measurements performed in the experiment in Sec.~\ref{sec:measurements}. This is followed by analyzing different post-selection schemes in Sec.~\ref{sec:PS-energy} and \ref{Sec:role:Diag}. 

\subsection{\label{sec:stating_problem}Stating the problem using laser driven HHG}

Recent advances in the quantum optical formulation of the process of HHG has indicated that coherent state driving fields are mapped to coherent output states due to the charge current induced by the strong laser field~\cite{lewenstein2021generation, cruz2024quantum}. 
This assumption holds when dipole moment correlations of the driven electron can be neglected~\cite{stammer_theory_2022} and when ground state depletion can be ignored~\cite{stammer2024entanglement}.
Typically this is given in HHG experiments using gas targets and moderate driving laser intensities, while for HHG in solid state systems~\cite{theidel_evidence_2024}, quantum correlated atomic systems~\cite{pizzi2023light}, or for higher driving intensities~\cite{stammer2024entanglement} deviations from the coherent state mapping are observed.
We note that the assumption that coherent states map to coherent states is also used when considering non-classical driving fields, such as bright squeezed vacuum, in which the field is decomposed into coherent states~\cite{gorlach_high-harmonic_2023} such that each contribution is considered independently.
In the following we will set the stage for the post-selection experiment using the HHG process and highlight its applicability with the product coherent state structure of the final field state after HHG. 

\begin{figure}
    \centering
	\includegraphics[width=1\columnwidth]{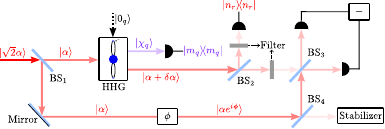}
	\caption{Schematic representation of the experimental setup including all relevant field modes. A laser source delivers radiation represented by the coherent state $\ket{\sqrt{2}\alpha}$, which is separated at a beam-splitter (BS). The lower arm is used as a local oscillator (LO) with variable phase $\phi$, while the upper arm induces the non-linear process of HHG. There, the input field $\ket{\alpha}$ is depleted leading to the state $\ket{\alpha + \delta \alpha}$, while the input vacuum of the harmonics results in a coherent state $\ket{\chi_q}$. The harmonic modes are measured, as well as the reflected part of the IR field after HHG, leading to projections on $\dyad{m_q}$ and $\dyad{n_r}$, respectively. These two measurements are used to generate the shot by shot correlation map as indicated in Fig.~\ref{fig:correlation_map}. The transmitted mode of the IR after the HHG passes to a homodyne measurement configuration where it overlaps with the LO $\ket{\alpha e^{\phi}}$ to reconstruct the quantum state of the IR field. The final BS in the LO arm is used to stabilize the intensity of the driving field eliminating intensity fluctuations by post-selection of those laser shots such that the standard deviation of the photon number $\sigma_n \equiv \sqrt{\Delta n}$ is in the range of $\sigma_n / \expval{n} \approx 0.5 \%$ and therefore minimizes classical intensity fluctuations associated with the shot-to-shot instabilities of the laser system. Further, there are two neutral density filters (NDF) placed in both IR arms such that the photon number is reduced to be compatible with the photodiodes. Note that the NDF does not significantly alter the photon number fluctuations, which remain around $\sigma_n / \expval{n} \approx 1.5 \%$ after the NDF.
    }
    \label{fig:setup}
\end{figure}

In more detail, we consider a laser source providing coherent radiation described by $\ket*{\sqrt{2} \alpha}$, where a $50:50$ beam-splitter (BS) separates the beam into a mode driving the process of HHG and a reference local-oscillator (LO) mode with variable phase $\phi$ for quantum state tomography (see Fig.~\ref{fig:setup} for a sketch of the experiment). 
Hence, the field interacting with the atomic HHG medium for the generation of high-order harmonics is described by the coherent state $\ket{\alpha}$ while the harmonic field modes are initially in the vacuum $\bigotimes_q \ket{0_q}$. 
The non-linear electron dynamics induced by the intense driving field leads to an electron charge current such that the outgoing field after HHG is described by (a detailed microscopic derivation can be found in~\cite{stammer_quantum_2023})
\begin{align}
    \ket{\alpha + \delta \alpha} \bigotimes_q \ket{\chi_q},
\end{align}
where the fundamental mode experienced a shift $\delta \alpha$ taking into account the depletion of the driving field and the harmonic fields modes are given by coherent states with amplitude $\chi_q$. 
We find that the final field state after the HHG process is in a product state, implying that measurements on one mode do not affect the other field modes. 
However, there is a crucial detail missing in this description: the fact that the depletion of the fundamental driving field depends on the harmonic amplitudes, i.e. $\delta \alpha = \delta \alpha (\chi_q)$, which makes the state of the IR driving field correlated with the harmonic XUV modes. 
Furthermore, since HHG is a parametric process in which the electron after the interaction returns to the initial state, the entire energy transfer is solely between the different field modes. 
It is now a matter of isolating the process of HHG from unwanted secondary effects such as excitation or ionization.


\subsection{\label{sec:measurements}Measurements in the post-selection experiment}

\begin{figure}
    \centering
	\includegraphics[width=1\columnwidth]{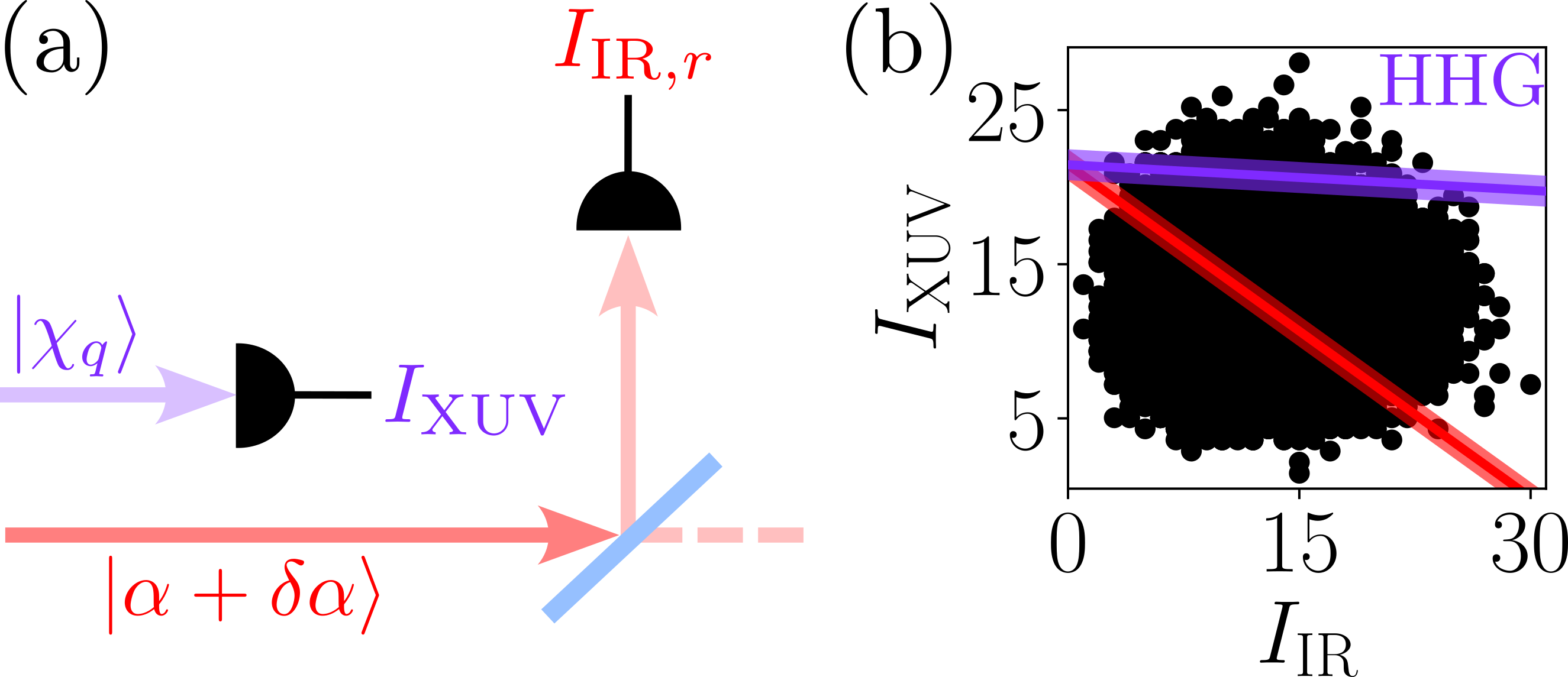}
	\caption{Correlation map of the measured IR vs. XUV shot by shot intensity distribution. (a) Zoom into the experimental setup indicating the reflected IR mode and the measured XUV mode. (b) Pictorial correlation map of the shot by shot XUV and IR intensity. Since the experiment is measuring the photocurrent, the measured signal is proportional to the photon number. We thus show $I_{\text{XUV}} = \abs{\chi_q}^2$ and $I_{\text{IR}} = \abs{\alpha + \delta \alpha}^2$.
    Thus, the anti-correlation diagonal (in red) corresponds to the photon number conserving events, while the diagonal (in purple) on the energy conserving events from Eq.~\eqref{Eq:Energy:All:modes} when considering the 13th harmonic order, and therefore condition the field on the HHG process. In this correlation map, we set $\alpha = 25$, $\delta\alpha = -15$ and $\chi_q = \abs{\delta \alpha}/\sqrt{q}$. The $y$-axis has been normalized to the mean value of the $I_{\text{IR}}$ distribution, i.e., multiplying the $y$-axes by $\bar{I}_{\text{IR}}/\bar{I}_{\text{XUV}}$, as per Ref.~\cite{tsatrafyllis2017high}.
    }
    \label{fig:correlation_map}
\end{figure}

In the previous section we have introduced the experimental setup, while in this section we discuss the post-selection (PS) scheme in more detail. After the process of HHG the IR driving field is given by the shifted coherent state $\ket{\alpha + \delta \alpha}$ and the harmonics are in a coherent state $\ket{\chi_q}$. The IR field passes through a $50:50$ BS such that the state is separated into a transmitted ($t$) and reflected ($r$) mode 
\begin{align}
\label{eq:state_afterHHG}
    \ket*{(\alpha + \delta \alpha)/\sqrt{2}}_t \otimes \ket*{(\alpha + \delta \alpha)/\sqrt{2}}_r \otimes \ket{ \{\chi_q \} }, 
\end{align}
where $\ket{ \{ \chi_q \} } \equiv \bigotimes_q \ket{\chi_q}$ is a short-hand notation for all harmonic field modes.
Then, measurements on the reflected mode and the XUV modes are performed.
These measurements are done through photodetectors measuring a photocurrent. The photocurrents are proportional to the field intensity, i.e. the photon number of each mode, such that in each shot $(i)$ we have measured
\begin{align}
\label{eq:operator_noPS1}
    \hat{\mathcal{O}} (n_r, \{ m_q \} ) = \dyad{n_r(i)} \bigotimes_q \dyad{m_q(i)},    
\end{align}
where $\{ m_q \}$ is the collection of all photon numbers $m_q$ of the harmonics.
The total operation on all three field modes is therefore 
\begin{align}
\label{eq:operator_noPS2}
    \mathds{1}_t \otimes \hat{\mathcal{O}} (n_r, \{ m_q \}),    
\end{align}
where $\mathds{1}_t = \sum_{n_t} \dyad{n_t}  $ is the identity on the transmitted mode.
The measured signal of the reflected IR mode and the XUV modes are fluctuating from shot to shot and the measured intensities can be correlated, as shown in Fig.~\ref{fig:correlation_map}~(b) in which the XUV intensity is shown for the IR intensity measured in each shot. This is due to the sampling of photon number states from the coherent state distribution in both field modes. 
We emphasize that the correlation map in Fig.~\ref{fig:correlation_map}~(b) is different from those observed experimentally~\cite{lewenstein2021generation, rivera-dean_strong_2022}. While the correlation map shown here has already excluded effects such as ionization or excitation of the electron when solving for the final field state used to obtain the final field state, while these effects are of course present in the experiment. Therefore, the statistics shown in Fig.~\ref{fig:correlation_map}~(b) originate from the intrinsic photon number fluctuations of a coherent state, whereas in the experiment all processes are present. We thus expect that the experimental fluctuations leading to the distribution are larger than the intrinsic fluctuations of a coherent state. 

However, we are interested in conditioning the outgoing quantum state of the IR field on the process of high harmonic generation. Due to energy conservation we know that the IR driving field is depleted during the HHG interaction such that an increased harmonic intensity is correlated with a decreased IR intensity after the interaction~\cite{tsatrafyllis2017high, stammer_energy_2024}. 
This is due to the parametric nature of the HHG process in which the electron returns to its initial state after the interaction. Consequently, no energy is stored in the matter system, and it is only responsible for mediating the energy between the field modes. We note that during the light-matter interaction, the electron can be excited from the ground state to another bound state or the continuum, and it is thus an experimental task to remove these unwanted events. 

We are now interested to relate the measured photon numbers to the energy conserving events of HHG by means of post-selection techniques.  

\subsection{\label{sec:PS-energy}Post-selection according to energy conservation}

Based on the measurement in Eq.~\eqref{eq:operator_noPS1} we consider the post-selection (PS) case, in which the photons lost in the IR driving field are correlated to the harmonic photons such that energy is conserved~\cite{stammer_energy_2024}.
The condition of shot-to-shot energy conservation for the photons in the field is given by 
\begin{align}
    \label{Eq:Energ:cons}
    n_t(i) + n_r(i) + \sum_q q m_q(i) = n_0,
\end{align}
where $n_t$, $n_r$ and $m_q$ are the photon numbers in the transmitted, reflected and XUV modes, respectively, and $n_0 = \abs{\alpha}^2$ is the initial state boundary condition before the HHG interaction. Note that this expression is given in units of the driving laser frequency $\omega$.
In the energy conservation relation we have separated the IR field into a reflected and transmitted mode, due to the 50:50 $\operatorname{BS}_2$ (see Fig.~\ref{fig:setup}), such that the relation with the XUV modes involves both parts of the IR field. 
Since the IR photon number is not changed by $\operatorname{BS}_2$ the energy conservation still holds, even though only the reflected mode is measured.

In other words, Eq.~\eqref{Eq:Energ:cons} implies that for the collection of generated harmonic photons $m_q$ the number of $\sum_q q m_q = \Delta n_{IR}$ photons from the IR field are absorbed.
This essentially allows to post-select on those events in which $\Delta n_{IR}$ number of IR photons are removed from the initial driving field.
Consequently, in order to take into account the energy conservation in the post-selection scheme, the condition of the IR photons in the transmitted mode for each shot is given by 
\begin{align}\label{Eq:Energy:All:modes}
    n_t(i) = n_0 - \Delta n_{IR}(i) - n_r(i).  
\end{align}

This ensures that the energy conservation is properly accounted for, while measuring the photon number $n_r$ and $m_q$ in the reflected and XUV mode, respectively.  
Therefore, when post-selection is performed on the energy conserving events in Eq.~\eqref{Eq:Energy:All:modes}, we introduce the post-selection (PS) operation 
\begin{equation}
\label{eq:final_PS_operation}
    \hat{\mathcal{O}}_{PS} = \sum_{n_r + \Delta n_{IR} = n_0/2} \delta_{PS} \, \dyad*{n_t (i)} \otimes  \hat{\mathcal{O}}(n_r, \{ m_q \} ).
\end{equation}

The sum goes over the energy conserving diagonal in which $n_r + \Delta n_{IR} = n_0/2$, and $\delta_{PS} = \delta_{n_t,n_0/2}$ ensures that the energy conservation is considered over all modes. 
Due to the energy conservation based on exact photon numbers, we find that the post-selection operator in \eqref{eq:final_PS_operation} is diagonal in the Fock basis, i.e. resulting in a Fock state of the transmitted mode. This is intuitive for the exact energy conservation case, since a given number of photons in the reflected IR mode and the harmonic mode result in a well defined photon number for the transmitted mode. 
The crucial argument of total energy conservation between all field modes provides the insights to generate non-classical field states in the sampled measurement data, and is motivated from the fact that the displacement of the driving field $\delta \alpha( \{ \chi_q \})$ is the a function of the harmonic amplitudes $\{ \chi_q \} $. The number of photons generated in the harmonic modes corresponds to the number of photons subtracted from the initial driving field, and counting the harmonic photons allows to infer on the number of subtracted photons~\cite{stammer_energy_2024}.

However, we note that, strictly speaking, energy conservation is not given based on exact photon number counts because the initial coherent state of the light field is not an eigenstate of the free-field Hamiltonian, and therefore does not have a vanishing dispersion in energy or photon number. But energy is, of course, conserved on the level of the Hamiltonian operator. 
Therefore, and instead of considering strict conservation of energy on the basis of photon numbers, we take into account the fluctuations of the photon numbers of the field modes and replace 
\begin{align}\label{Eq:fuzzy:energy}
    \delta_{PS} & \to \exp[- \frac{(n_t - n_0 + \Delta n_{IR} + n_r)^2}{2 \sigma^2}], 
\end{align}
where we set $\sigma^2 \approx n_0/2$, though the effect of varying values of $\sigma^2$ is considered more deeply in Sec.~\ref{Sec:role:Diag}. This is in fact an adaptation of the absence of knowledge about the precise underlying photon number of the driving field, and therefore the energy conserving subspace of the interaction is not well defined~\cite{stammer_energy_2024}.

\begin{figure}
    \centering
	\includegraphics[width=1\columnwidth]{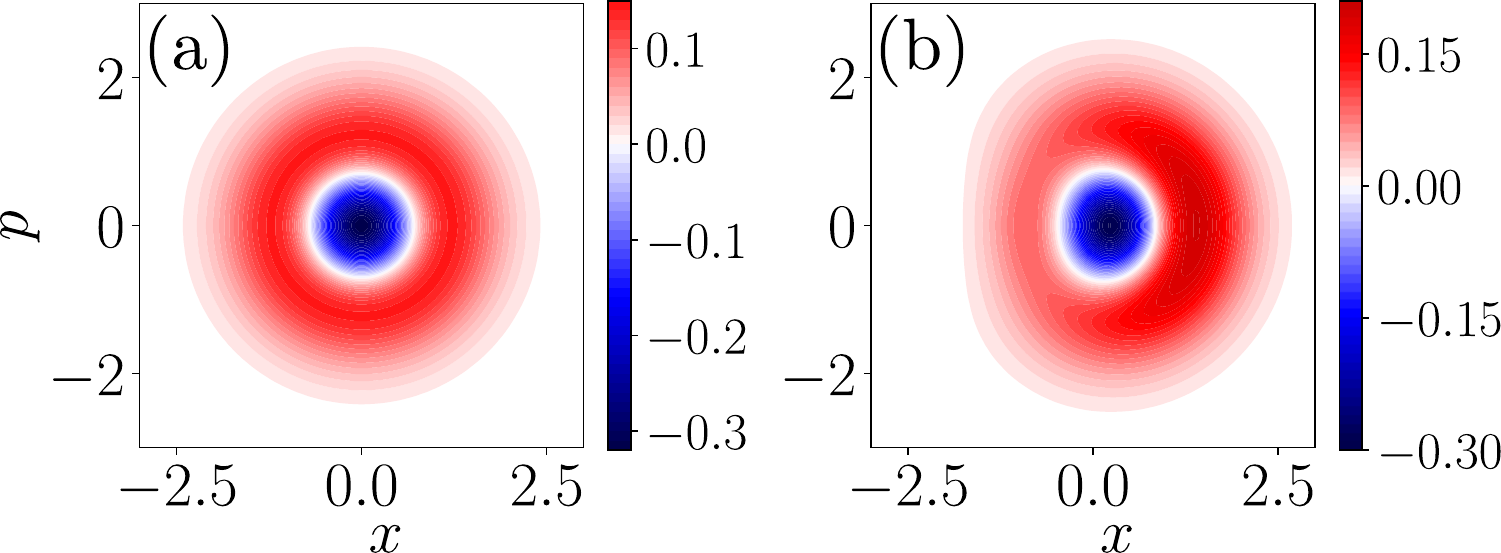}
	\caption{In (a), the postselected state when considering exact energy conservation in Eq.~\eqref{Eq:Energy:All:modes}. In (b), when considering the fuzzy version in Eq.~\eqref{Eq:fuzzy:energy}. The parameters have been set to $\alpha = 1.2$, $\delta\alpha = -0.3$, $q_1 = 13$ and $q_2 = 15$. In this figure, the $x$ and $p$ axes represent the optical quadratures, with $x \equiv \expval{\hat{x}} = \expval{\hat{a}+\hat{a}^\dagger}/\sqrt{2}$ and $p \equiv \expval{\hat{p}} = \expval{\hat{a}-\hat{a}^\dagger}/(i\sqrt{2})$.}
    \label{fig:PS_state_numerical}
\end{figure}

Given the description of the post-selection scheme and operators, we shall now look at the state of the transmitted mode given by 
\begin{align}
\label{Eq:postselected:state}
    \hat \rho_t = \Tr_{r,q}[\hat{\mathcal{O}}_{PS} \, \hat \rho \, \hat{\mathcal{O}}_{PS}^\dagger],
\end{align}
where the state, on which this PS is performed is given by $\hat \rho = \dyad{\Psi}$, where $\ket{\Psi}$ is the state in Eq.~\eqref{eq:state_afterHHG} after the HHG interaction and the BS$_2$. Numerical evaluation of the PS operation in Eq.~\eqref{Eq:postselected:state} is displayed in Fig.~\ref{fig:PS_state_numerical}, where the Wigner function of the light field in the transmitted mode is shown. 
Due to the numerical implementation we restrict the analysis to two harmonic modes, $q_{1,2} = 13, \ 15$ and sample the data points with the PS constraint from Eq.~\eqref{eq:final_PS_operation}. 
We find that the associated Wigner function in Fig.~\ref{fig:PS_state_numerical} displays clear negativities when the strict energy conservation based on the exact photon numbers in Eq.~\eqref{Eq:Energy:All:modes} is considered. Even the fuzzy version of the energy conservation in Eq.~\eqref{Eq:fuzzy:energy} shows the same negativities and only the symmetry of the Wigner function gets slightly distorted. 

The Wigner functions of the single mode transmitted field shown in this section are obtained by direct computation of the Wigner function from the state in Eq.~\eqref{Eq:postselected:state}. The PS constrained is included in the numerical evaluation by expressing the states in the Fock basis and only considering those elements which satisfy the PS condition.

\subsection{Post-selection without energy conservation}\label{Sec:role:Diag}

\subsubsection {Using a single harmonic mode}

In Eq.~\eqref{Eq:Energy:All:modes}, we considered a postselection on the photon number of the reflected mode satisfying an energy conservation relation. However, the presented postselection scheme allows for evaluating different diagonals, which do not necessarily need to satisfy the energy conservation relation.~Here, we evaluate the impact of choosing different diagonals on the characteristics of the postselected state through its influence on the resulting Wigner function. For simplicity, throughout this section, we restrict the analysis to a single-mode scenario, rewriting Eq.~\eqref{Eq:Energy:All:modes} as
\begin{equation}\label{Eq:Diff:Diags}
    n_r(i) + \kappa\ m_q(i) = c,
\end{equation}
where $\kappa$ is treated as a tunable parameter, while the value of $c$ remains as detailed in Eq.~\eqref{Eq:Energy:All:modes}. It is worth noting that $\kappa = 1$ corresponds to the scenario where post-selection is performed on the anti-correlation diagonal of the intensity distribution corresponding to conservation of photon number and not energy. 
While $\kappa = q$ corresponds to the desired energy conservation as in Eq.~\eqref{Eq:Energy:All:modes}.
The nature of the post-selection scheme allows to process the data post measurement such that different diagonals can be chosen.

\begin{figure}
    \centering
    \includegraphics[width = 1\columnwidth]{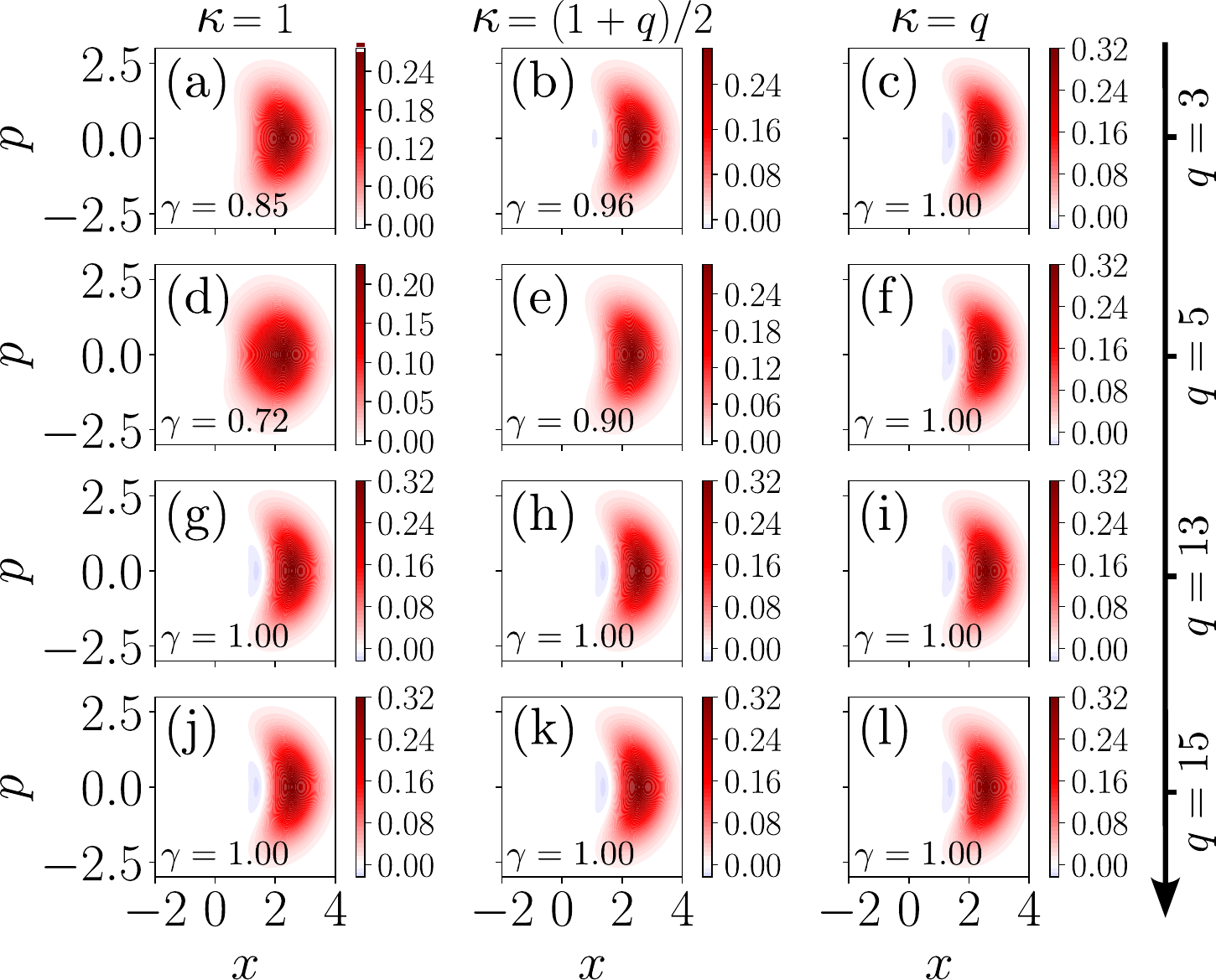}
    \caption{Wigner functions obtained by considering different diagonals (determined through the value of $\kappa$) when postselecting on the transmitted modes. Here, we have set $\alpha = 3.0$ and $\delta \alpha = -1.0$. In the first row we set $q=3$, in the second $q=5$, in the third $q=13$ and in the fifth $q=15$. The purity of the postselected state, $\gamma = \tr(\hat{\rho}_t^2)$, is shown in each of the panels. 
    } 
    \label{Fig:Diff:Diag}
\end{figure}

The results from this analysis are shown in Fig.~\ref{Fig:Diff:Diag}.~From top to bottom, we set $q=3$, $q=5$, $q=13$ and $q=15$.~In all cases, we set $\alpha = 3.0$ and $\delta \alpha = -1.0$. Each row displays different values of $\kappa$, increasing from left to right, as indicated at the top of each panel.~For the two lowest harmonic orders, we observe that the value of $\kappa$ greatly influences the features of the Wigner function. While the Wigner function exhibits a non-Gaussian behavior, the negative regions are absent when $\kappa=1$ (panels (a) and (d)) and become more pronounced as $\kappa$ increases (panels (b), (c) and (f)), reaching a minimum value of $W(x^*,p^*) = -0.018$. Furthermore, for this harmonic order, the purity of the postselected state, evaluated as $\gamma = \tr(\hat{\rho}_{t}^2)$ and explicitly shown in each subplot, increases as the Wigner function decreases.~These two behaviors align with Hudson's theorem~\cite{hudson_when_1974}, which states that pure states have non-negative Wigner functions if and only if they are Gaussian.~Conversely, for the two highest harmonic orders shown in the second row, the features of the Wigner function and the purity of the postselected state are more resilient to the specific value of $\kappa$.~This resilience stems from the energy conservation postulate introduced in Eq.~\eqref{Eq:Energy:All:modes}, which determines the values of $n_t$ that significantly contribute to the postselection operator.~Thus, as the harmonic order $q$ increases, the support of $\hat{\mathcal{O}}_{PS}$ diminishes, thereby reducing the influence of $\kappa$.

\begin{figure}
    \centering
    \includegraphics[width=1\columnwidth]{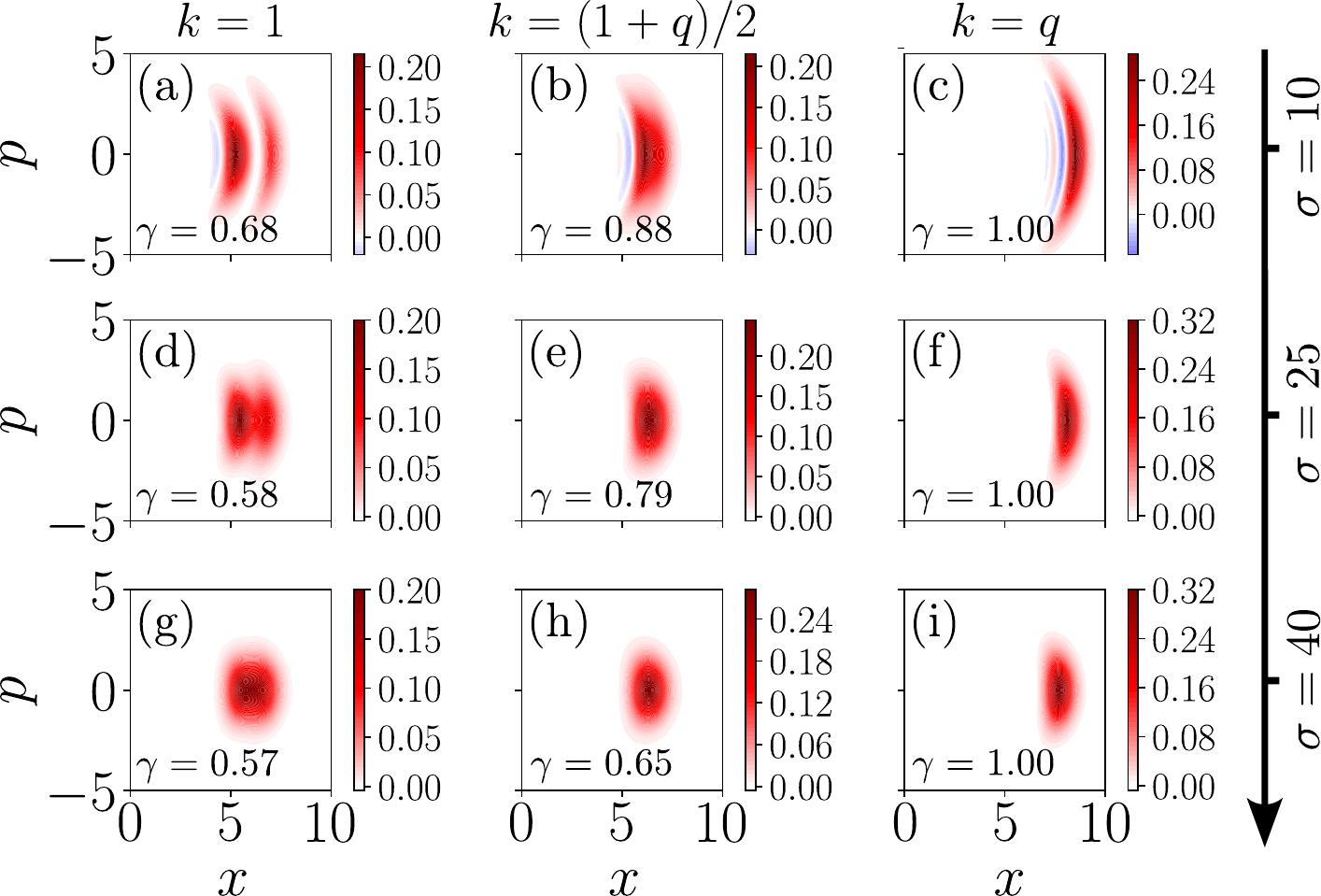}
    \caption{Wigner functions obtained by considering different diagonals when postselecting on the transmitted modes. Here, we have set $\alpha = 9.0$ and $\delta \alpha = -3.0$, while keeping $q=15$. Each row considers a different value of $\sigma$, i.e., the width of the considered diagonal. The purity of the postselected state, $\gamma = \tr(\hat{\rho}_t^2)$, is shown in each of the panels.}
    \label{fig:High:photon:diag}
\end{figure}

However, when considering high-harmonic orders, it is possible to increase the support of $\hat{\mathcal{O}}_{PS}$ on the state by considering larger values of $\alpha$. In Fig.~\ref{fig:High:photon:diag}, we examine the case where $\alpha = 9.0$ and $\delta \alpha = -3.0$---triple the values used in Fig.~\ref{Fig:Diff:Diag}---while fixing the harmonic order at $q=15$. Each column in this figure displays different diagonal configurations. Similar to the results obtained in the low harmonic order regime in Fig.~\ref{Fig:Diff:Diag}, we observe that the Wigner negativities and the purity of the state increase as $\kappa$ changes from 1 to $q$. Unlike Fig.~\ref{Fig:Diff:Diag}, however, each row considers a different value of $\sigma$, i.e., the width of the diagonal. This parameter also plays an important role in determining both the purity and the presence of Wigner negativities. Specifically, for the latter, the greater the width of the diagonal, the less pronounced the Wigner negativities become. This is consistent with the fact that increasing $\sigma$ enlarges the support of $\hat{\mathcal{O}}_{PS}$ on the state, such that in the limit $\sigma \to \infty$, the postselection operator approaches the identity, leading to Gaussian Wigner functions.

\subsubsection{Using multiple harmonic modes}

While the previous subsection focused on the analysis of individual modes, we can generalize Eq.~\eqref{Eq:Diff:Diags} to include an arbitrary number of harmonic modes. Specifically, we can express it as
\begin{equation}
    n_r(i) + \sum_{q}\kappa_q m_q(i) = c,
\end{equation}
which allows for a greater degree of tunability when selecting the correlated photon numbers.

\begin{figure}
    \centering
    \includegraphics[width=1\columnwidth]{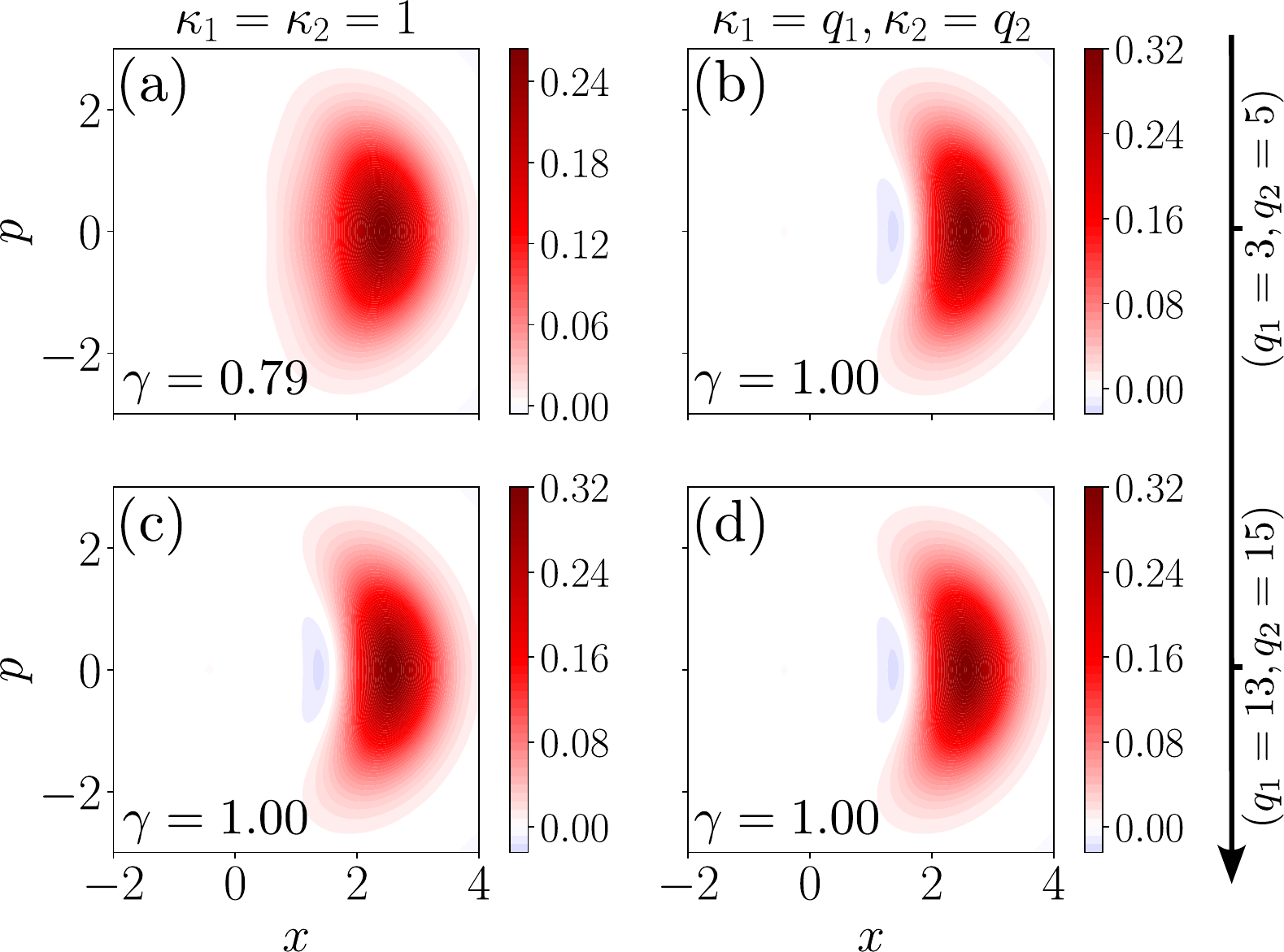}
    \caption{Wigner functions obtained by considering different diagonals (determined through the values of $\kappa_1$ and $\kappa_2$) when postselecting on the transmitted modes.  Here, we have set $\alpha = 3.0$ and $\delta \alpha = -1.0$. In the first row we set $(q_1=3,q_2=5)$, while in the second row $(q_1=13,q_2=15)$. The purity of the postselected state, $\gamma = \tr(\hat{\rho}_t^2)$, is shown in each of the panels.  
    }
    \label{Fig:Diff:Diag:2:harms}
\end{figure}

In Fig.~\ref{Fig:Diff:Diag:2:harms}, we present the results obtained when selecting a pair of harmonic modes $(q_1,q_2)$, with the diagonals fixed to $\kappa_1 = \kappa_2 = 1$, for the first column; while $\kappa_1=q_1$ and $\kappa_2 = q_2$ for the second one. The observed trend closely resembles that seen in Fig.~\ref{Fig:Diff:Diag}, interestingly showing squeezing-like features though with non-Gaussian characteristics for the cases where $\gamma = 1.00$, revealed through the presence of Wigner negativities. For the lowest harmonic order pair with $\kappa_1=\kappa_2 = 1$ (panel (a)), the resulting Wigner function lacks from negative regions but still exhibits a non-Gaussian behaviour, with the postselected state showing a purity below one. When $\kappa_1$ and $\kappa_2$ are increased to respectively match the corresponding harmonic orders, non-classical features on the obtained Wigner functions become more prominent, and the purity approaches one (panel (b)). Conversely, for the highest harmonic orders, the resulting features become independent of the values of $\kappa_1$ and $\kappa_2$ (panels (c) and (d)), with purities equal to one and Wigner functions exhibiting negative regions. 

\section{\label{sec:comparison}Comparison with optical cat states}

The theoretical models describing the post-selection on HHG experiment that have been used thus far were successful in reproducing the experimental observations~\cite{lewenstein2021generation,rivera-dean_strong_2022,stammer_high_2022,stammer_theory_2022,stammer_quantum_2023}, even though they rely on phenomenological approaches.
In essence, the key idea of these models is to project the state of the driving field onto the subspace that has been altered by the HHG process. Finding this subspace was the delicate task of the previous approaches. While the first experimental realization of post-selection on HHG~\cite{lewenstein2021generation} was described using operators projecting the state $\ket{\alpha + \delta \alpha}$ on the subspace ``orthogonal'' to the initial state, i.e. $\Pi_{HHG} = \mathds{1} - \dyad{\alpha}$. Note that the projector $\Pi_{HHG}$ is not orthogonal to the initial state $\ket{\alpha}$ due to the overcompleteness of the coherent state basis. 
The resulting state $\ket{\alpha + \delta\alpha} \to \ket{\psi} = (\mathbbm{1} - \dyad{\alpha})\ket{\alpha+\delta\alpha}$ showed good agreement with the experimental data obtained from the HHG conditioning experiment. 
Further advancing this description, in Refs.~\cite{stammer_high_2022, stammer_theory_2022} the correlations between the coherent state amplitudes $\delta \alpha = \delta \alpha (\{ \chi_q \} )$ have been introduced via a set of wavepacket modes corresponding to joint HHG excitations of all modes. Using the formalism of the quantum theory of measurement the positive operator valued measure (POVM) for the HHG process was rigorously introduced~\cite{stammer_high_2022, stammer_theory_2022}, and found the same projector $\Pi_{HHG}$ in the limit of many harmonic modes participating in the process. 
All these approaches leading to the same projector result in a superposition between two coherent states, explicitly given by
\begin{equation}\label{Eq:Old:Cat}
    \ket{\psi}
        = \ket{\alpha + \delta\alpha}
        - \braket{0}{\delta\alpha}
            \ket{\alpha}.
\end{equation}

In this section, we undertake a theoretical comparison between the state defined above, and the postselected state introduced in Eq.~\eqref{Eq:postselected:state} through numerical sampling. Following this comparison, we simulate experimental outcomes by performing numerical statistical sampling, which mimics the experimental shots carried out in homodyne detection experiments under ideal conditions. Finally, we compare these simulated results with the experimental data reported in Ref.~\cite{rivera-dean_strong_2022}.


To estimate how well the conditioning approach in Eq.~\eqref{Eq:Old:Cat} compares to the postselected state in Eq.~\eqref{Eq:postselected:state}, we use the fidelity $\mathcal{F}$ as our figure of merit. Specifically, we denote $\ket{\psi(\beta,\delta\beta)} = (1/N)(\ket{\beta + \delta \beta} - \xi \ket{\beta}$, with $\xi = \braket{0}{\delta \beta}$, and write the fidelity as $\mathcal{F}(\beta,\delta\beta)\equiv \langle \psi(\beta,\delta\beta)\vert \hat{\rho}_t\vert \psi(\beta,\delta\beta)\rangle$. We then compare both states by considering the following optimization problem
\begin{equation}
    \mathcal{F}(\beta^*,\delta\beta^*)
        = \max_{\beta,\delta\beta} \mathcal{F}(\beta,\delta\beta),
\end{equation}
which we solve using brute force search approaches by defining a grid of size $100 \times 100$, with elements within the range $\beta,\delta\beta \in [10^{-3}, 3]$. It is worth noting that, in this optimization, we consider $\delta\beta > 0$ and allow regimes where $\delta\beta \geq \beta$ to properly align the orientation of the Wigner functions obtained by Eq.~\eqref{Eq:postselected:state} with those resulting from models based on the form of Eq.~\eqref{Eq:Old:Cat}.

\begin{figure}
    \centering
    \includegraphics[width = 1 \columnwidth]{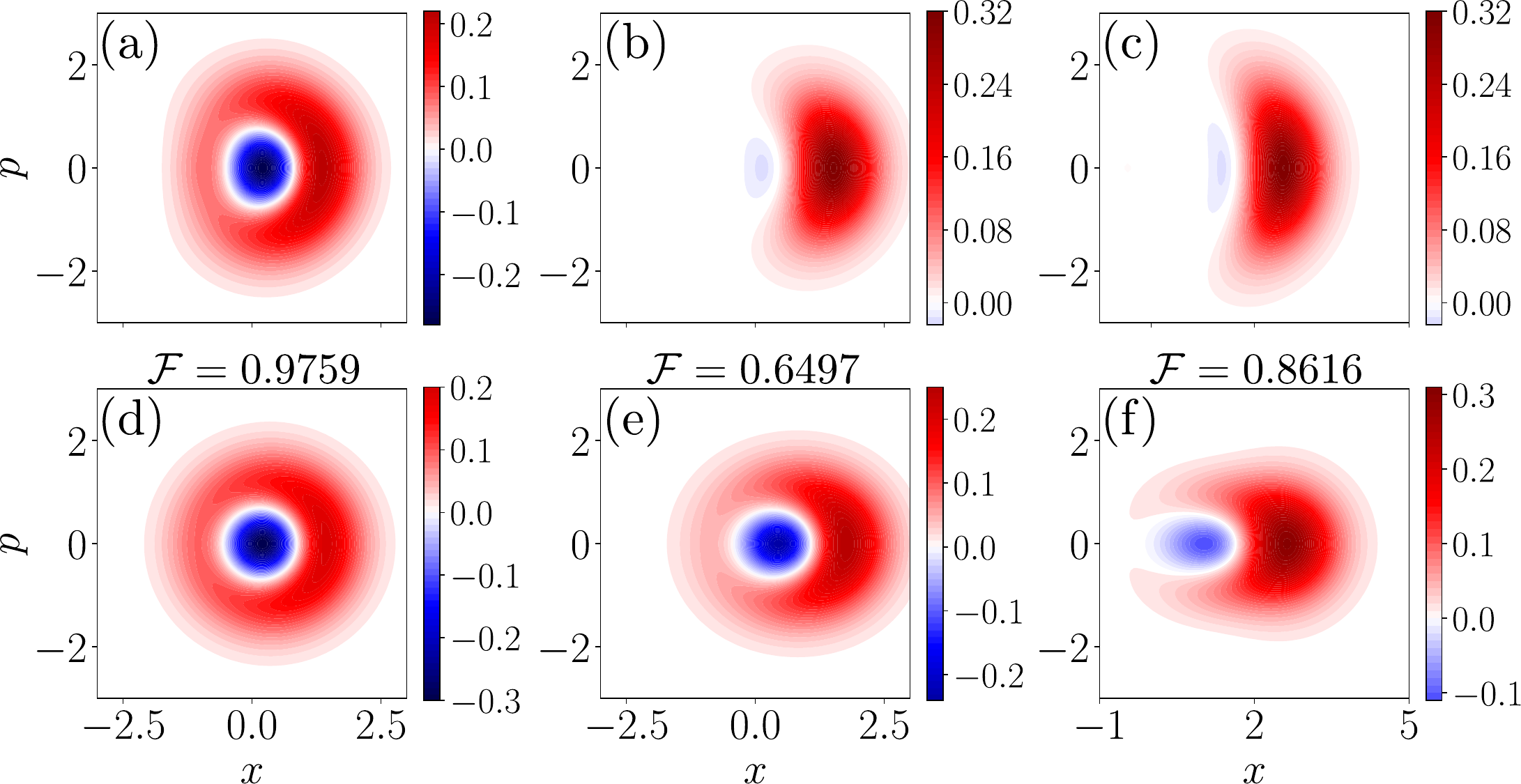}
    \caption{Comparison of the postselected state in Eq.~\eqref{Eq:postselected:state} with that in Eq.~\eqref{Eq:Old:Cat} by means of fidelity optimizations. The first row shows the postselected state for different values of $(\alpha,\delta\alpha)$, in particular (a) $(1.2, -0.3)$, (b) $(2.0,-0.7)$ and (c) $(3.0,-1.0,)$. The second row shows the optimal state $\ket{\psi(\beta^*,\delta\beta^*)}$, with parameters $(\beta,\delta\beta)$ given by $(10^{-3},0.455)$ in (d), $(10^{-3},0.91)$ in (e) and $(0.1,1.70)$ in (f).  
    }
    \label{Fig:Fidelity}
\end{figure}

The results of this optimization are shown in Fig.~\ref{Fig:Fidelity} for different values of $\alpha$ and $\delta\alpha$ (see caption). In particular, the first row presents the Wigner function of Eq.~\eqref{Eq:postselected:state} for the considered parameters, while the second row displays the same phase-space distribution for the optimal state $\ket{\psi(\beta^*,\delta\beta^*)}$. For the three cases considered here, the optimal values of the fidelity are 97.59\%, 64.97\% and 86.16\%, from left to right, respectively. This indicates that the theoretical model proposed in Ref.~\cite{lewenstein2021generation} performs particularly well in the very low-photon number regime, as is the case of panels~(a) and (d). For intermediate low-photon numbers, corresponding to those of panels~(b) and (d), the fidelity gets reduced mainly due to an overestimation of the quantum superposition presented by Eq.~\eqref{Eq:Old:Cat},  leading to larger negative regions in the Wigner function. As the mean photon number of the postselected state increases, the fidelity also increases, as illustrated in panels~(e)-(f), though not as much as in panels~(a)-(d), resulting in reasonably high fidelity values.

\section{\label{sec:limitations}Finite-size sampling experimental and their limitations}\label{Sec:Exp:Lim}

In this section we provide a detailed tour into the tomographic reconstruction of quantum states of light. We consider a homodyne configuration which allows to reconstruct the Wigner function of the measured field. 
Since the post-selection on HHG experiment is a sampling experiment we consider its shot by shot nature of the realistic measurement, the limitations of the reconstruction method and analyze fluctuations in the driving field. 

So far the Wigner functions shown in the manuscript are obtained by directly using the quantum state of the field mode under investigation. Now, in order to study experimental imperfection, we perform a numerical quantum sampling experiment~\cite{royer_wigner_1977,SchleichBookCh3}. This allows to take into account the finite nature of every sampling experiment, in order to understand the limitations of the post-selection protocol. While finite-size sampling plays a crucial role in all (quantum) experiments, it was for instance studied in quantum metrology~\cite{meyer2023quantum}, state discrimination~\cite{audenaert2012quantum}, efficient quantum measurement~\cite{crawford2021efficient} or quantum cryptography~\cite{yao2022quantum}.


\subsection{Comparison with existing experimental data}\label{Sec:Comp:Exp}

Experimental reconstructions of the Wigner function rely on homodyne detection measurements~\cite{smithey_measurement_1993} in which the optical state we want to characterize is combined with a well-characterized coherent state with a controllable phase, typically referred to as the local oscillator (LO), using a 50 : 50 beam splitter, as shown at the right of Fig.~\ref{fig:setup}. The intensities of the two output modes of the beam splitter are then measured and subtracted, leaving us with
\begin{equation}\label{Eq:Homodyne}
    \Delta I
        = I_{\text{in}} - I_{\text{LO}}
        \propto 
        \langle 
            \hat{E}_{\text{in}}(t) \hat{E}_{\text{LO}}(t)
        \rangle,
\end{equation}
with $\hat{E}(t) \equiv \hat{a}e^{-i\omega t}+ \hat{a}^\dagger e^{i\omega t}$ the electric field operator. Given that the state of the local oscillator is $\lvert\abs{\alpha}e^{i\phi}\rangle$, and after integrating over a given number of optical cycles, we can rewrite \eqref{Eq:Homodyne} as
\begin{equation}
    \Delta I 
        \propto 
            \langle 
                \hat{a}_{\text{in}}e^{i\phi} 
                + \hat{a}^\dagger_{\text{in}}e^{-i\phi}
            \rangle 
        \equiv \langle \hat{x}_{\text{in}}(\phi) \rangle,
\end{equation}
that is, we obtain the average value of photonic quadratures along the direction determined by $\phi$. In experimental reconstructions, this quantity is approximated by
\begin{equation}
    \langle \hat{x}_{\text{in}}(\phi) \rangle
        \cong
            \dfrac{1}{N_{\text{shots}}}
                \sum^{N_{\text{shots}}}_{i=1}
                    x^{(i)}_{\text{in}}(\phi),
\end{equation}
where $x^{(i)}_{\text{in}}(\phi)$ correspond to single shot measurements of the quadrature $\hat{x}_{\text{in}}(\phi)$, and $N_{\text{shots}}$ represents the total number of shots for a specific value of $\phi$.

A potential way for reconstructing the Wigner function from homodyne measurements is by means of the inverse Radon transformation~\cite{leonhardt_measuring_1997,lvovsky_continuous-variable_2009}
\begin{equation}\label{Eq:inv:Radon}
    W_{\text{exp}}(x,p)
        \cong \dfrac{1}{2\pi^2 N_{\phi}}
            \sum_{m=1}^{N_{\phi}}
                K(z_m),
\end{equation}
with $z_m \equiv x \cos(\phi_m) + p \sin(\phi_m) - \langle \hat{x}_{\text{in}}(\phi_m)\rangle$, $N_{\phi}$ the total number of angles considered and $K(z_m)$ the so-called integration kernel, given by $K(z) = \frac12 \int^{\infty}_{-\infty} \dd \xi \abs{\xi}\exp[i\xi z]$. However, one of the main problems with this kernel function is that it is infinite at $z=0$. Consequently, in numerical implementations, the integration limits are substituted by finite ones, $\pm k_c$, with $k_c$ chosen to reduce numerical artifacts related to the reconstruction while maintaining the features displayed by the Wigner function~\cite{lvovsky_continuous-variable_2009}. More specifically, in the reconstruction performed in Ref.~\cite{lewenstein2021generation} it was observed that $k_c \sim 4$ was a good enough value to reduce the numerical artifacts while maintaining a well-resolved Wigner function. In our case, we instead find that $k_c = 2$ provides us with better Wigner function reconstructions, as it will be discussed in Section~\ref{Sec:Exp:Lim}.

\begin{figure}
    \centering
    \includegraphics[width = 1\columnwidth]{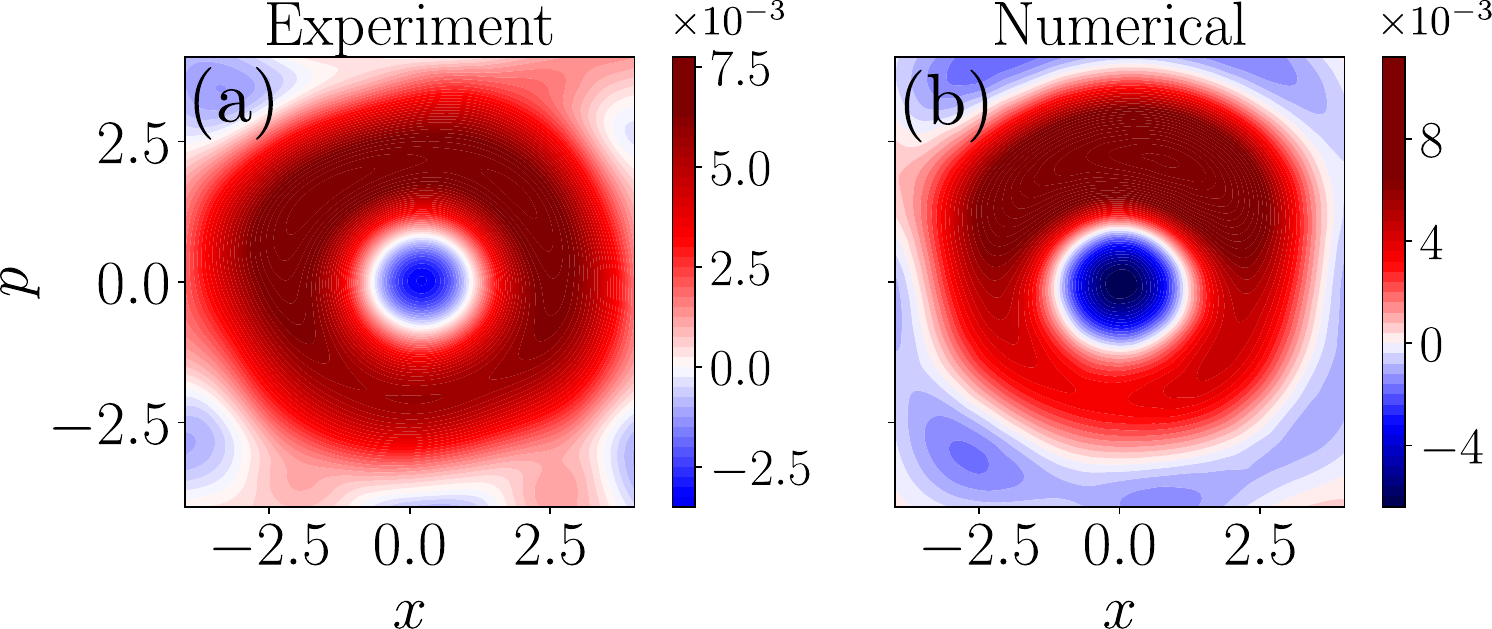}
    \caption{Comparison between experimental data and the numerical approach. In (a), we show the experimental Wigner function obtained in Ref.~\cite{rivera-dean_strong_2022}. In (b), the one obtained from the numerical sampling when setting $\alpha = 1.2$, $\delta\alpha = -0.3$, $N_{\text{shots}} = 100$, $N_{\phi} = 20$ and $k_c = 2$. The $x$ and $p$ axes of the experimental data were aligned to ensure that the main features of both Wigner functions are captured within the same phase space span.}
    \label{Fig:com:exp:theory}
\end{figure}

To do a proper comparison between the experimental and theoretical data, we consider a statistical sampling from the probability distribution $P(x)$ resulting from measuring $\hat{x}_{\text{in}}(\phi)$ in the postselected state given in \eqref{Eq:postselected:state} (see Appendix~\ref{App:sampling} for details). This allows us to obtain a set of $N_{\text{shots}}\times N_{\phi}$ pairs $(\phi^{(i)}_m, x^{(i)}_{\text{in}}(\phi^{(i)}_m))$, with the outcomes $x^{(i)}_{\text{in}}(\phi^{(i)}_m)$ distributed according to $P(x)$, that simulate the outcomes of the experiment in Fig.~\ref{fig:setup} under ideal conditions, i.e., in the absence of any experimental sources of noise. Then, by means of Eq.~\eqref{Eq:inv:Radon}, we can reconstruct the Wigner function from the outcomes $(\phi^{(i)}_m, x^{(i)}_{\text{in}}(\phi^{(i)}_m))$. This numerical sampling of the Wigner function provides us with a better tool to perform fidelity estimations with respect to experimental data. Here, we compare the results obtained from the numerical sampling in the previous section with those obtained in Ref.~\cite{rivera-dean_strong_2022} (see Fig.~\ref{Fig:com:exp:theory}~(a)), where Wigner function negativities below the artifacts induced by the quantum tomography protocol were observed. To achieve this, we first identify values of $\alpha$ and $\delta\alpha$ that yield Wigner functions in good agreement with the experimental ones. In our case, these values are $\alpha = 1.2$ and $\delta\alpha = -0.3$.

Using this set of parameters, we followed the approach depicted earlier to perform the numerical sampling, and applied the inverse Radon transformation to recover a matrix $\mathsf{W}_{\text{num}}$ describing the numerical Wigner function presented in Fig.~\ref{Fig:com:exp:theory}~(b). For this, we used a total number of shots $N_{\text{shots}} = 100$, considered a total of $N_{\phi} = 20$ angles and set $k_c = 2$ (this choice is justified later on in Sec.~\ref{Sec:Exp:Lim}). The dimensions of this matrix were adjusted to match those of the experimental data, denoted hereupon as $\mathsf{W}_{\text{exp}}$. Additionally, the $x$ and $p$ axes of the experimental data were aligned to ensure that the main features of both Wigner functions in Fig.~\ref{Fig:com:exp:theory} are captured within the same phase space span. This alignment is crucial as it depends on the parameters used for the reconstruction, particularly $k_c$ as it will be discussed in Sec.~\ref{Sec:Exp:Lim} (see Fig.~\ref{Fig:sampling:diff:kc}). By doing this, we observe in Fig.~\ref{Fig:com:exp:theory} the same overall volcano-like shape for both the experimental and numerical Wigner functions, with a minimum located at the origin whose depth varies depending on the nature of the Wigner function, and with a maximum located on top of it.

Given that both $\mathsf{W}_{\text{exp}}$ and $\mathsf{W}_{\text{num}}$ have different normalizations, we first normalize them to a consistent scale. To do so, we define $\bar{\mathsf{W}}_{\text{num}}$ and $\bar{\mathsf{W}}_{\text{exp}}$ as $\bar{\mathsf{W}} = \mathsf{W}/\norm{\mathsf{W}}_F$, where $\norm{\mathsf{W}}_F$ denotes the Frobenius norm of matrix $\mathsf{W}$, $\norm{\mathsf{W}}_F = \sqrt{\sum_{i}\sum_j \abs{\mathsf{W}_{ij}}^2}$. This normalization ensures that $\norm{\bar{\mathsf{W}}_{\text{num}}}_F=\norm{\bar{\mathsf{W}}_{\text{exp}}}_F = 1$. By normalizing this way, for any two real matrices $\bar{\mathsf{A}}$ and $\bar{\mathsf{B}}$, we have $\langle \bar{\mathsf{A}},\bar{\mathsf{B}}\rangle_F \leq 1$, where $\langle \cdot, \cdot \rangle_F$ denotes the Frobenius inner product between two matrices. This allows us to compare the similarity between $\bar{\mathsf{W}}_{\text{num}}$ and $\bar{\mathsf{W}}_{\text{exp}}$. Following this approach, we obtain that $\langle \bar{\mathsf{W}}_{\text{num}},\bar{\mathsf{W}}_{\text{exp}}\rangle_F = 0.8898$, suggesting a fairly high degree of similarity between the experimental data and that obtained through the numerical sampling. As mentioned earlier, the numerical sampling considered here does not include any of the experimental nuances about the realistic postselection protocol, which we expect to further increase the similarity between theory and experiment, or can be seen as sources of noise in the experiment. 

\begin{table*}[t]  
\caption{Summary of the minimum of the respective Wigner functions $W_{\text{min}}$ and the visibility of the quantum interference $\abs{W_{\text{min}} / W_{\text{max}}}$ via the ratio of the Wigner minima and maxima for the different approaches to describe the post-selection experiment in HHG. The same parameters for the theoretical approaches as for maximizing the fidelity with the experimental data from Ref.~\cite{rivera-dean_strong_2022} Fig.~7(b) are used when finding the Wigner minima and interference visibility. For the theoretical models the values of the coherent state parameters are given in the respective brackets $[\alpha, \delta \alpha]$.   }
\label{table:features}
  \centering
\begin{tabular}{ |c||c|c| c | c | c|}
 \hline
 & Exp. \cite{lewenstein2021generation} & Exp. \cite{rivera-dean_strong_2022} Fig.7 (b) & Exp. \cite{rivera-dean_strong_2022} Fig. 8 & $\ket{\psi} = \ket{\alpha+\delta\alpha} - \braket{0}{\delta \alpha} \ket{\alpha}$ & Sampling (this work) \\
\hline
\hline 
$W_{\text{min}}$ & -0.01& -0.04& 0.02& -0.01 [$\alpha = 10^{-3}, \delta\alpha = 0.45$] &-0.03 [$\alpha = 1.2, \delta\alpha = -0.3$] \\
\hline
$\abs{W_{\text{min}} / W_{\text{max}}}$ & 0.05 &0.42 &0.17 & 0.63 [$\alpha = 10^{-3}, \delta\alpha = 0.45$]& 0.55 [$\alpha = 1.2, \delta\alpha = -0.3$] \\
\hline 
\end{tabular}
\end{table*}

\begin{table*}[t]  
\caption{Summary of the different approaches to describe the post-selection experiment in HHG by providing the fidelity $\mathcal{F}$ between the different cases. The fidelities between the experimental data and the two theoretical approaches, the analytical and the sampling approach, are maximized by varying the theoretical parameters, which are given in each entry alongside the fidelity. Specifically, we search for parameters in Eq.~\eqref{Eq:postselected:state} that best reproduce the experimentally measured mean photon number and yield similar features to the observed Wigner function. Following a similar optimization procedure as in Fig.~\ref{Fig:Fidelity}, we identify the optimal state of the form given in Eq.~\eqref{Eq:Old:Cat}. Using these two analytical states, we conduct numerical sampling as described in Sec.~\ref{Sec:Comp:Exp} and Appendix~\ref{App:sampling}, and perform a fidelity comparison similar to that in Fig.~\ref{Fig:com:exp:theory}. }
\label{table:fidelity}
  \centering
\begin{tabular}{ |c||c|c| }
 \hline
 &  $\ket{\psi} = \ket{\alpha+\delta\alpha} - \bra{0}\ket{\delta \alpha} \ket{\alpha}$ & Numerical sampling (this work) \\
\hline
\hline 
Exp. \cite{lewenstein2021generation} &  $\mathcal{F}= 0.54$, $[\alpha = 0.1, \delta \alpha = 0.92]$ & $\mathcal{F}= 0.65$, $[\alpha = 1.8, \delta \alpha = -0.5]$ \\
\hline
Exp. \cite{rivera-dean_strong_2022} Fig. 7 (b) & $\mathcal{F}=0.89$ $[\alpha = 10^{-3}, \delta \alpha = 0.45]$ & $\mathcal{F}=0.89$ $[\alpha = 1.2, \delta \alpha = -0.3]$   \\
\hline
Exp. \cite{rivera-dean_strong_2022} Fig. 8  & $\mathcal{F}=0.91$ $[\alpha = 1.2, \delta \alpha = 1.84]$ & $\mathcal{F}=0.93$ $[\alpha = 5.0, \delta \alpha = -1.6]$ \\
\hline
\end{tabular}
\end{table*}

Finally, for completeness, Tables~\ref{table:features} and \ref{table:fidelity} summarize a comparison of the Wigner function features and fidelity estimates, respectively, including the experimental results of Ref.~\cite{lewenstein2021generation,rivera-dean_strong_2022}. The approach described in this section has been followed, with additional details provided in the Caption of Table~\ref{table:fidelity}.

\subsection{Shot-to-shot measurement}

\begin{figure}[b]
    \centering
    \includegraphics[width = 1\columnwidth]{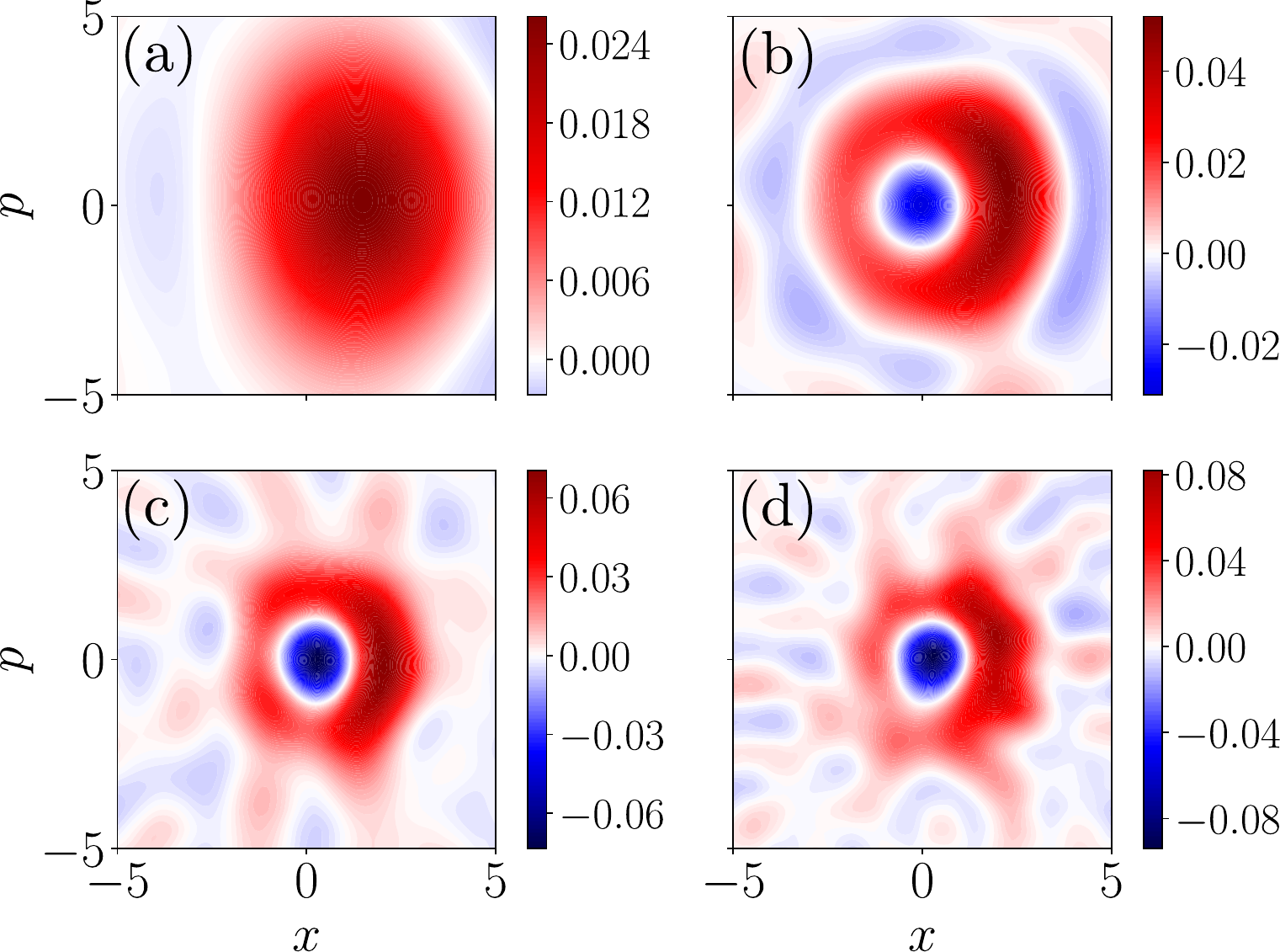}
    \caption{Reconstruction of the Wigner function from the numerical sampling for different values of $k_c$. More specifically, we have set (a) $k_c = 1$, (b) $k_c = 2$, (c) $k_c=3$ and (d) $k_c=4$. In all cases, we have set $N_{\text{shots}} = 100$, $N_{\phi} = 20$ uniformly spread from $[0, \pi]$, $\alpha = 1.2$ and $\delta\alpha = -0.3$. The exact Wigner function for the corresponding state is shown in Fig.~\ref{Fig:Fidelity}~(b).}
    \label{Fig:sampling:diff:kc}
\end{figure}

Thus far, we can distinguish three main limitations that could affect the experimental reconstruction of our Wigner function: the number of shots performed per angle ($N_{\text{shots}}$), the number of angles considered in the reconstruction ($N_{\phi}$), and the value of $k_c$. Here, we present how these different values affect, from a numerical perspective, the reconstruction of the Wigner functions shown in Fig.~\ref{Fig:Fidelity}. We begin by analyzing the influence of $k_c$ on the reconstructed Wigner functions. To do so, we fix $N_{\text{shots}} = 100$ and $N_{\phi} = 20$, leading to a total number of 2000 outcomes, and set $\alpha = 1.2$ and $\delta\alpha = -0.3$ such that the exact Wigner function corresponds to that in Fig.~\ref{Fig:Fidelity}~(a), depicting two maxima: a small one to the left of $(x=0,p=0)$ and a prominent one around ($x\simeq 2, p=0$). A minimum value is observed at $(x=0,p=0)$. The results from this analysis are presented in Fig.~\ref{Fig:sampling:diff:kc}. In all cases, we observe the presence of artifacts, i.e., extra negative (in blue) and positive (in red) regions compared to those appearing in the exact distribution. It is worth noting that, in some cases, the observed negative regions are comparable in magnitude to that of the central one at $(x=0,p=0)$.

Given that the generated data lacks additional experimental noises, these extra artifacts stem directly from the reconstruction method, as $K(z)$ involves the integral of an oscillatory function, with the number of oscillations considered depending on the value of $k_c$. In other words, a finite value of $k_c$ introduces a low-pass filter that attenuates the high-frequency components of our reconstruction. While increasing values of $k_c$ allow for the extraction of finer details about the Wigner function, they also introduce high-frequency artifacts. Hence, one can observe a trade-off on the values of $k_c$: the smaller the values, the worse the reconstruction of the Wigner function is, as shown in Fig.~\ref{Fig:sampling:diff:kc}~(a); the higher the values, the more resolved the Wigner function becomes at the expense of acquiring more artifacts in the reconstruction, as shown in  Fig.~\ref{Fig:sampling:diff:kc}~(d). For the values of $k_c$ considered here, we observe that $k_c = 2$ best fits these tradeoff requirements, justifying therefore its use in Fig.~\ref{Fig:com:exp:theory}. However, to the best of our knowledge, there is no established intuition for selecting this cutoff value a priori, beyond a trial-and-error approach, and bench-marked to a known distribution (most commonly a reconstruction of a Gaussian from a laser system providing a coherent state).

\begin{figure}
    \centering
    \includegraphics[width = 1\columnwidth]{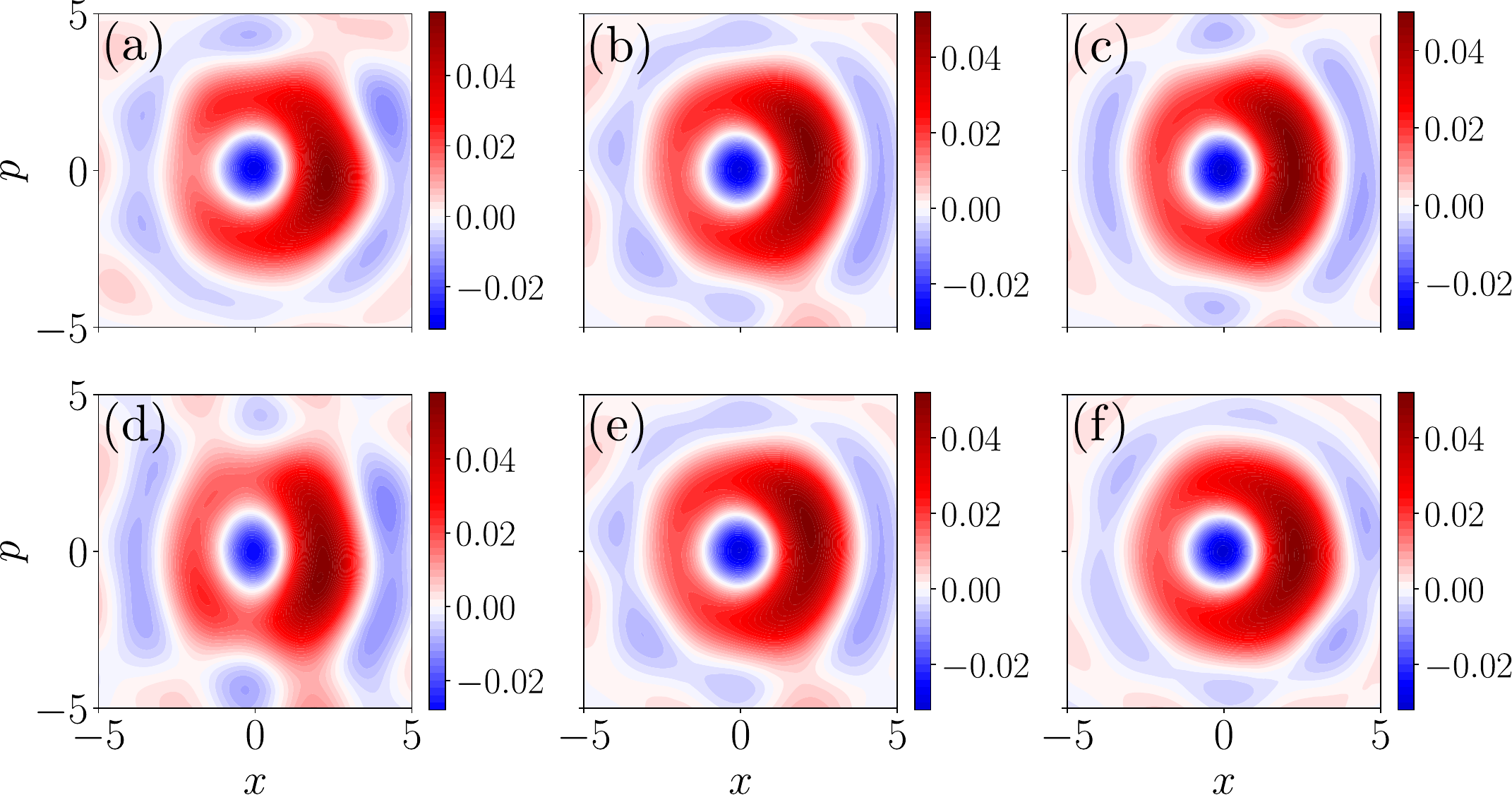}
    \caption{Reconstruction of the Wigner function from the numerical sampling for different values of $N_{\text{shot}}$ (first row) and $N_{\phi}$ (second row). More specifically, in the first row $N_{\phi} = 20$ while $N_{\text{shots}} = 50$ in (a), $N_{\text{shots}} = 100$ in (b) and $N_{\text{shots}} = 500$ in (c). Conversely, in the second tow $N_{\text{shots}} = 100$ while $N_{\phi} = 10$ in (d), $N_{\phi} = 20$ in (e) and $N_{\phi} = 50$ in (f). In all plots, $k_c = 2$, $\alpha = 1.2$ and $\delta\alpha = -0.3$.}
    \label{Fig:sampling:diff:shots}
\end{figure}

Under the same conditions for the postselected state as those chosen in Fig.~\ref{Fig:sampling:diff:kc}, in Fig.~\ref{Fig:sampling:diff:shots} we instead examine the influence of varying $N_{\text{shots}}$ (first row) and of $N_{\phi}$. More specifically, in Fig.~\ref{Fig:sampling:diff:shots}~(a) to (c), we fixed $N_{\phi} = 20$ and varied the value of $N_{\text{shots}}$ (50, 100 and 500 respectively). Alternatively, in Fig.~\ref{Fig:sampling:diff:shots}~(d) to (f), we fixed $N_{\text{shots}} = 100$ and varied the value of $N_{\phi}$ (10, 20 and 50 respectively). In all panels, we have fixed $k_c = 2$. As observed, an increase in either of these sampling points provides enhanced resolution of the measured Wigner functions. For instance, compared to Fig.~\ref{Fig:Fidelity}~(a), the region located at the left of the origin becomes better resolved. Nevertheless, this increase in the amount of data does not reduce the presence of the artifacts mentioned earlier, which instead require the use of more elaborate reconstruction methods in order to be mitigated~\cite{lvovsky_continuous-variable_2009,hradil_3_2004}.

To summarize, in Sec. \ref{sec:comparison} and \ref{sec:limitations},  we have evaluated the agreement of the analytical expression of the optical ``cat'' state give by Eq.~\eqref{Eq:Old:Cat} with the state reconstructed using the numerical sampling approach, and with existing experimental data. The more important values are those from Table \ref{table:fidelity}, in which the numerical sampling approach developed in this work is compared to the experimental data from Refs.~\cite{lewenstein2021generation, rivera-dean_strong_2022}. We do this to be as close as possible to the experiment, via simulating the finite nature of the shot-to-shot experiment, and analyzing the influence of the finite statistics on the homodyne trace and Wigner function via inverse Radon transformation. In Table \ref{table:fidelity} the values of fidelity in comparison to the experimental data from Ref.~\cite{lewenstein2021generation} and those from Ref.~\cite{rivera-dean_strong_2022} show $F = 0.65$ and $F \ge 0.9$, respectively. This difference is associated to experimental imperfections, and  the implementation of the post-selection technique which has been improved in Ref.~\cite{rivera-dean_strong_2022} compared to Ref.~\cite{lewenstein2021generation}. This explains why the fidelity is smaller for the ``cat'' state measured in Ref.~\cite{lewenstein2021generation} compared to the ``cat'' states measured in Ref.~\cite{rivera-dean_strong_2022}. In other words, we conclude that the comparison shows that the analytical expression of the ``cat'' state shown in Eq.~\eqref{Eq:Old:Cat} captures the overall features of the post-selected state, and in particular provides a good factual representation of the state.

\subsection{Intensity fluctuations of the driving field}

In addition to potential noise contributions from the quantum tomography reconstruction and the experimental specifications associated with the applied measurements, additional sources of infidelity can arise from the employed laser source. Here, we consider the impact of classical fluctuations in the laser field amplitude. 
For a perfect coherent state driving field $\ket{\alpha}$ with amplitudes necessary to drive the HHG process on the order of $\abs{\alpha} \sim 10^6$ the average photon number is on the order of $\expval{n} = \abs{\alpha}^2 \sim 10^{12} $. For the coherent state the variance of the photon number is given by $\Delta n = \abs{\alpha}^2 \sim 10^{12}$, such that the standard deviation of the photon number measurement for such intense coherent states is on the order of $\sigma_n = \sqrt{\Delta n} \sim 10^6$. Therefore, the quantum fluctuations due to the Poisson distribution of the coherent state can not be resolved for such high intensity laser sources due to $\sigma_n / \expval{n} = 1/\abs{\alpha} \sim 10^{-6}$, and classical fluctuations of the driving laser system dominate. 
As outlined in the experimental setup in Fig.~\ref{fig:setup} a stabilization measurement is performed which selects only those events in which the intensity of the laser shot is within the range of $\sigma_n / \expval{n} \approx 0.5 \%$.
Even the presence of the neutral density filter (NDF), needed to reduce the photon numbers for the tomographic reconstruction, does not significantly alters these fluctuations which remain at $\sigma_n / \expval{n} \approx 1.5 \%$ after the NDF.
Consequently, instead of delivering a coherent state with a well-defined amplitude $\abs{\alpha}$, the state generated by the driving field source corresponds to
\begin{equation}\label{Eq:intens:fluc}
    \hat{\rho}
        = \int
            \dd \abs{\alpha}
                p(\abs{\alpha})
                \dyad{\alpha},
\end{equation}
where $p(\abs{\alpha}) \propto \exp[-(\abs{\alpha} - \alpha_0)/(2\Tilde{\sigma}^2)]$, with $\Tilde{\sigma}\neq \sigma$ in \eqref{Eq:fuzzy:energy}. In the following, we restrict ourselves to relatively small values of $\Tilde{\sigma}$ so that we can assume that $\delta\alpha$ remains approximately the same despite the varying values of $\alpha$ present in each laser shot.

\begin{figure}
    \centering
    \includegraphics[width=1\columnwidth]{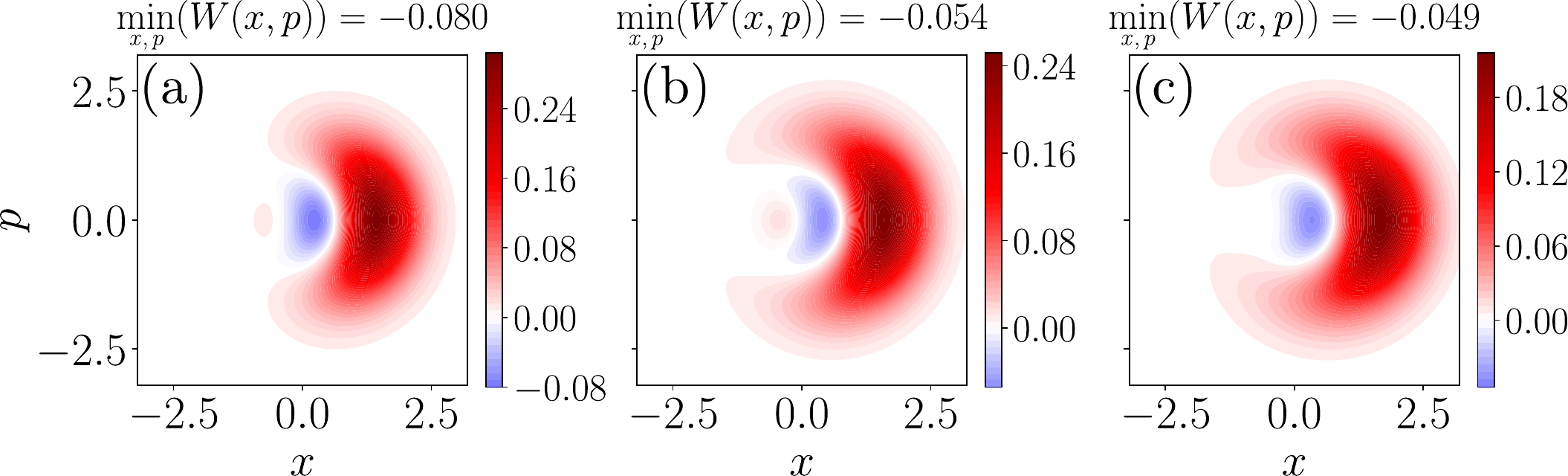}
    \caption{Wigner functions obtained when considering Eq.~\eqref{Eq:intens:fluc} as the initial state of the driving field, for different values of $\Tilde{\sigma}$. Specifically, in (a) $\Tilde{\sigma} = 0.02$, (b) $\Tilde{\sigma} = 0.22$ and (c) $\Tilde{\sigma} = 0.33$. Here, we have fixed $\alpha_0 = 2$ and $\delta\alpha = -0.3$.}
    \label{Fig:fluct:intens}
\end{figure}

In Fig.~\ref{Fig:fluct:intens}, we present the results obtained for increasing values of $\Tilde{\sigma}$ from left to right, respectively. It is worth noting that, despite the presence of the intensity fluctuations, the results display non-negative regions in their Wigner function representation (specified at the top of each subplot), while the overall features in terms of shape remain the same. However, these negativities get reduced as $\Tilde{\sigma}$ increases. This suggests that, despite the presence of intensity fluctuations in the considered laser source, these do not constitute a major source of infidelity in the generated state. Nevertheless, in conjunction with other factors such as those studied earlier, intensity fluctuations can contribute to the absence of clear negative regions, which in some cases could become comparable to the artifacts introduced by quantum tomography methods.

\subsection{Detector efficiency}

In the experiments performed in Refs.~\cite{lewenstein2021generation, rivera-dean_strong_2022} photodetectors are used to measure the photon numbers of the field for the correlations map as presented in Fig.~\ref{fig:correlation_map}~(a). However, photodetectors are not perfect and not every photon generated is also detected (in addition to photon losses in the experimental setup). 
Therefore, the photon number of the $q$-th harmonic generated in the HHG experiment in laser shot $(i)$ is given by $M_q(i)$, while the measured photon number is different $m_q(i) = \eta_q M_q(i)$, where $\eta_q \leq 1 $ and takes into account photon losses in the setup and the detector efficiency. 
Assuming that the detector efficiency is the same for each harmonic mode, we write $\eta_q = \eta$, which for the photodetectors used in the experiment in Refs.~\cite{lewenstein2021generation, rivera-dean_strong_2022} is around $\eta = 0.2$, i.e. approximately $20\%$ detector efficiency. 
Thus, for the argument of the energy conservation diagonal in Eq.~\eqref{Eq:Energy:All:modes} we can see that it modifies to 
\begin{align}
    n_r(i) + \sum_q \eta \,  q \, M_q(i) = n_0 / 2.
\end{align}

This shows, that the energy conserving diagonal for ideal photodetection is changed in such a way that the slope of the diagonal is reduced by the factor $\eta$ and tends towards the total photon number conserving slope in which $q = 1$.

\section{\label{sec:conclusion}Conclusions}

In this work we have provided a detailed analysis of the post-selection schemes recently introduced for the process of HHG (although the scheme is independent of the precise process as long as energy conservation in the light generation process is given, e.g.~for all parametric processes). 
We have outlined the key ingredients of the scheme, which is the treatment of energy conservation from the measurement of the harmonic photon number and a portion of the driving field after the interaction. Knowing the photon number from the measurement of the photocurrent in photodiodes allows to infer on the energy conversion between the modes, and using that for the generation of $m_q$ photons of the harmonic mode $q$, the number of $q \, m_q$ photons from the IR field are converted. 
Post-selection on these events allows to condition the IR state on the photon subtracted initial coherent state resembling a Wigner function of a coherent state superposition, i.e. an optical cat state.  

We have shown that using intensity measurements and classical post-processing allows to generate measurement statistics which are characteristic to those of non-classical states of light, while having only classical input states. The measurement characteristic looks like \textit{as if} there would be a non-classical light field generated. 
The classical post-selection is performed after the measurement such that the non-classical states is only generated upon the detection including the state tomography measurement. 
However, it is important to note that it is only the measurement statistics which is important. Every quantum experiment is a sampling experiment, and here, the sampling reproduces the same statistics \textit{as if} there is an optical cat-state propagating in the experiment. This is further supported by the observation of non-classical features in the experimentally reconstructed Wigner function despite intentionally including elements such as the neutral density filters which would otherwise destroy the non-classicality. Such cat states have been shown to be able to drive non-linear optical phenomena~\cite{lamprou_nonlinear_2023}, and are particularly robust against imperfections in quantum metrology~\cite{stammer2024metrological}.
Nevertheless, this scheme opens the way for novel experiments towards the generation of new quantum states of light. It was explicitly shown that different post-selection conditions lead to different quantum states which allows for further control of the final measurement statistics. 
With this work we open the perspective for post-selection schemes and conditioning approaches in strong field processes, and shows the richness of the quantum state engineering protocol. 

Finally, we emphasize that the theoretical description of the quantum state of light after HHG is still intensively discussed and investigated. For instance, the approximate expression of the coherent state maps to coherent states, and the underlying assumptions leading to such a mapping are missing a clear presentation and the regime of its validity is still not known. 
We believe there is a deeper underlying disparity between two pictures, in this case a wave-like description for the quantum state of the field after HHG and a particle-like measurement in the conditioning and post-selection approach. While the product coherent state output of the field after HHG is based on an oscillating charge current, the energy conservation of the HHG process is not reflected in the coherent state amplitudes. In contrast, the photon picture of absorption of IR photons for the generation of XUV photons directly involve the energy conservation of the process, which is not apparent in the coherent state description. 
To conclude, we emphasize that the non-Gaussianity of the generated quantum states comes from the post-selection of the measurement statistics, which is evidently non-Gaussian. This leads to non-Gaussian states albeit the input states are Gaussian.

\begin{acknowledgments}

P.S. acknowledges funding from the European Union’s Horizon 2020 research and innovation programme under the Marie Skłodowska-Curie grant agreement No 847517. 
E.P. acknowledges Royal Society fellowship funding under URF\textbackslash R1\textbackslash 211390.
M.F.C. acknowledges financial support from the Guangdong Province Science and Technology Major Project (Future functional materials under extreme conditions - 2021B0301030005) and the Guangdong Natural Science Foundation (General Program project No. 2023A1515010871).
ICFO group acknowledges support from:
Europea Research Council AdG NOQIA; 
MCIN/AEI (PGC2018-0910.13039/501100011033, CEX2019-000910-S/10.13039/501100011033, Plan National FIDEUA PID2019-106901GB-I00, Plan National STAMEENA PID2022-139099NB, I00, project funded by MCIN/AEI/10.13039/501100011033 and by the “European Union NextGenerationEU/PRTR" (PRTR-C17.I1), FPI); QUANTERA MAQS PCI2019-111828-2); QUANTERA DYNAMITE PCI2022-132919, QuantERA II Programme co-funded by European Union’s Horizon 2020 program under Grant Agreement No 101017733);
Ministry for Digital Transformation and of Civil Service of the Spanish Government through the QUANTUM ENIA project call - Quantum Spain project, and by the European Union through the Recovery, Transformation and Resilience Plan - NextGenerationEU within the framework of the Digital Spain 2026 Agenda;
Fundació Cellex;
Fundació Mir-Puig; 
Generalitat de Catalunya (European Social Fund FEDER and CERCA program, AGAUR Grant No. 2021 SGR 01452, QuantumCAT \ U16-011424, co-funded by ERDF Operational Program of Catalonia 2014-2020); 
Barcelona Supercomputing Center MareNostrum (FI-2023-3-0024); 
Funded by the European Union. Views and opinions expressed are however those of the author(s) only and do not necessarily reflect those of the European Union, European Commission, European Climate, Infrastructure and Environment Executive Agency (CINEA), or any other granting authority.  Neither the European Union nor any granting authority can be held responsible for them (HORIZON-CL4-2022-QUANTUM-02-SGA  PASQuanS2.1, 101113690, EU Horizon 2020 FET-OPEN OPTOlogic, Grant No 899794),  EU Horizon Europe Program (This project has received funding from the European Union’s Horizon Europe research and innovation program under grant agreement No 101080086 NeQSTGrant Agreement 101080086 — NeQST); 
ICFO Internal “QuantumGaudi” project; 
European Union’s Horizon 2020 program under the Marie Sklodowska-Curie grant agreement No 847648;  
“La Caixa” Junior Leaders fellowships, La Caixa” Foundation (ID 100010434): CF/BQ/PR23/11980043;the European Union's HORIZON MSCA-2023-DN-01 project QU-ATTO under the Marie Skłodowska-Curie Grant Agreement No. 101168628.
P.Tzallas' group at FORTH acknowledges support from the Hellenic Foundation for Research and Innovation (HFRI) and the General Secretariat for Research and Technology (GSRT) under Grant Agreement CO2toO2 No. 015922, LASERLABEUROPE V (H2020-EU.1.4.1.2 Grant No. 871124), the European Union's HORIZON MSCA-2023-DN-01 project QU-ATTO under the Marie Skłodowska-Curie Grant Agreement No. 101168628, the LASERLABEUROPE V (H2020-EU.1.4.1.2 Grant No. 871124).

\end{acknowledgments}

\bibliography{references}{}

\newpage
\appendix

\begin{center}
    \textbf{APPENDIX}
\end{center}
\section{Numerical analysis of the homodyne statistical sampling}\label{App:sampling}
The numerical implementation used for the statistical sampling and subsequent Wigner function reconstruction through the inverse Radon transformation, as shown in Figs.~\ref{Fig:sampling:diff:kc}, \ref{Fig:sampling:diff:shots} and \ref{Fig:com:exp:theory}~(b), was entirely performed in Python by using built-in functions from different packages. Generally, the implementation involves numerically expressing the quantum optical state $\hat{\rho}$ with respect to a chosen basis. In our case we chose the Fock basis, as per standard of the \texttt{QuTiP} package~\cite{johansson_qutip_2012,johansson_qutip_2013}. Given that this basis naturally corresponds to an infinite-dimensional Hilbert space, it is necessary to impose a cutoff on the dimension to make the numerical analysis feasible. This cutoff must be high enough to ensure accurate representation of the state, and for the coherent state amplitudes considered in our study, $n_{\text{cutoff}}\in [100,200]$ suffices. Although higher values can be used, they come at the cost of increased computational time and memory resources.

Given the numerical representation of our state $\hat{\rho}$, we perform a homodyne measurement over it represented by the operator $\hat{I}(\phi) = \hat{a}_{\text{in}} e^{i\phi} + \hat{a}_{\text{in}}^\dagger e^{-i\phi}$. The set of eigenvalues $\{\lambda_i(\phi)\}$ of this operator determines the possible outcomes of the measurement, with the associated eigenstates $\{\ket{\varphi_i(\phi)}\}$ representing the state onto which the input state is projected after the measurement. Consequently, the set $\{\!\mel{\varphi_i(\phi)}{\hat{\rho}}{\varphi_i(\phi)}\}$ represents the probability $p(\lambda_i(\phi))$ of obtaining each outcome $\lambda_i(\phi)$. Both the sets $\{\lambda_i(\phi)\}$ and $\{p(\lambda_i(\phi))\}$ can be easily accessed using the \texttt{measurement\_statistics} function from the \texttt{measurements} subpackage of \texttt{QuTiP}. To simulate the outcomes of an experiment implementing homodyne detection on an input state $\hat{\rho}$, we need to numerically sample from the set of outcomes according to the specified probability distribution. This can be achieved using the \texttt{choices} function of the \texttt{random} package~\cite{python-random} which, given these two sets, can sample a specified number of times, corresponding to the number of experimental shots.

\begin{figure}[h!]
    \centering
    \includegraphics[width = 1\columnwidth]{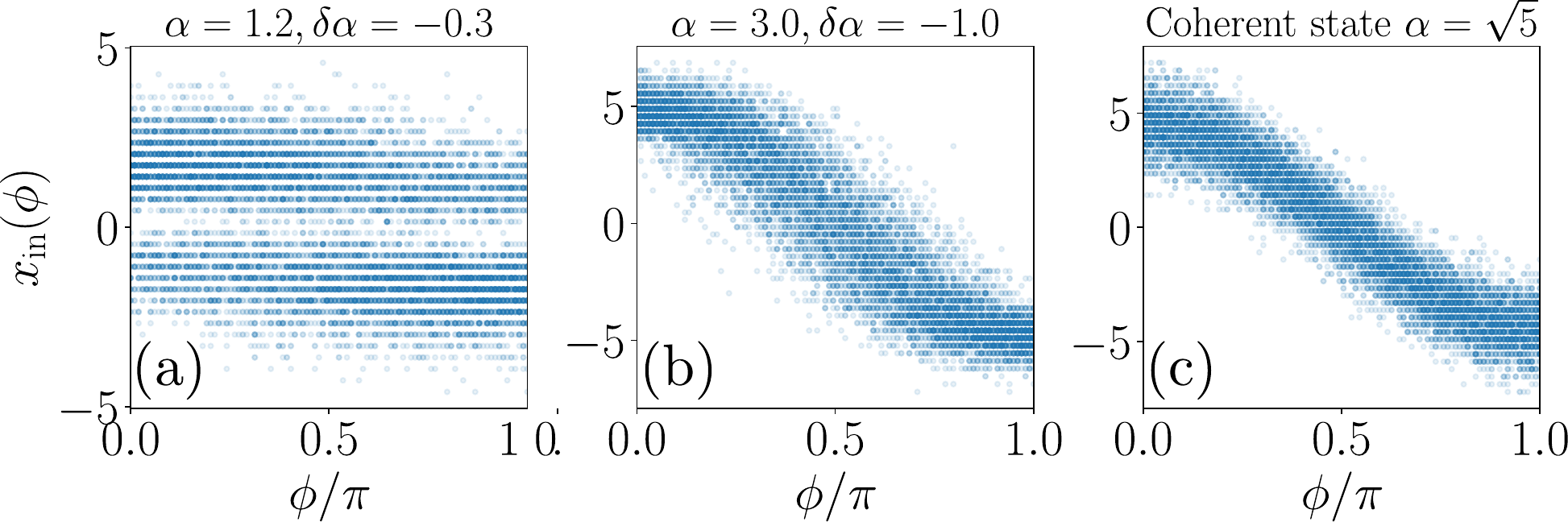}
    \caption{Example of numerically sampled homodyne trace used for the Wigner function reconstruction through the inverse Radon transformation. The parameters used for the postselected state from which the homodyne trace was obtained are $\alpha = 1.2$ and $\delta \alpha = -0.3$ for panel (a), matching Fig.~\ref{Fig:Fidelity}~(a); and $\alpha = 3.0$ and $\delta \alpha = -1.0$ for panel (b), matching Fig.~\ref{Fig:Fidelity}~(c). In panel (c), we considered a coherent state of amplitude $\alpha = \sqrt{5}$. For representational purposes, we have used $N_\phi = 100$ and $N_{\text{shots}} = 100$.  }
    \label{Fig:App:Homodyne}
\end{figure}

In Fig.~\ref{Fig:App:Homodyne} we present three examples of the results obtained for different states (see caption). For representational purposes, we chose $N_{\phi} = 100$ and $N_{\text{shots}} = 100$ (per angle). These represent the homodyne traces accessible experimentally through homodyne measurements. As discussed in the main text, these traces are subsequently post-processed using the inverse Radon transformation to approximately reconstruct the Wigner function. This step involves numerically integrating the integration kernel $K(z)$ in Eq.~\eqref{Eq:inv:Radon}, with appropriately chosen integration limits $\pm k_c$. The integration was performed using the \texttt{quad} function of the \texttt{Scipy} package~\cite{2020SciPy-NMeth}, with the integration parameters carefully adjusted to achieve convergence. Specifically, an upper bound of 1000 subintervals was employed in the adaptive algorithm. 

At the end of this process, we obtained a matrix $\mathsf{W}$, where each element $\mathsf{W}_{i,j}$ represents the value of the Wigner function at a specific point in phase-space. In our case, these matrices where  $20\times 20$ in size. To enhance the resolution of the resulting plots, we performed 2D interpolation of the data using the \texttt{griddata} function of the \texttt{interpolate} subpackage in \texttt{Scipy}.~Specifically, cubic polynomial interpolation was used.

\end{document}